\pdfoutput = 1

\documentclass[journal]{IEEEtran}
\ifCLASSINFOpdf
\else
\fi
\usepackage{amsmath}
\usepackage{algorithmic}

\usepackage{array}

\usepackage{stfloats}
\usepackage{url}

\usepackage[table,xcdraw]{xcolor}

\usepackage{graphicx}
\usepackage{gensymb}
\usepackage[normalem]{ulem}
\useunder{\uline}{\ul}{}
\usepackage{booktabs}
\usepackage{multirow}
\usepackage{siunitx}
\usepackage{tabulary}
\usepackage{booktabs}
\usepackage{xcolor}
\usepackage{graphicx}
\usepackage{amsmath}
\usepackage{threeparttable}
\usepackage{enumitem}
\usepackage{longtable}
\usepackage{xcolor}
\usepackage{graphicx}
\usepackage{amsmath}
\usepackage{threeparttable}
\usepackage{enumitem}
\usepackage{listings}
\usepackage{subfigure}

\definecolor{darkspringgreen}{rgb}{0.09, 0.45, 0.27}

\begin{document}
%
\title{XpulpNN: Enabling Energy Efficient and Flexible Inference of Quantized Neural Network on RISC-V based IoT End Nodes}


\author{\IEEEauthorblockN{Angelo Garofalo\IEEEauthorrefmark{1},
Giuseppe Tagliavini \IEEEauthorrefmark{1},
Francesco Conti \IEEEauthorrefmark{1}, 
Luca Benini \IEEEauthorrefmark{1} \IEEEauthorrefmark{2}, and
Davide Rossi \IEEEauthorrefmark{1}
}

\IEEEauthorblockA{\IEEEauthorrefmark{1}Department of Electrical, Electronic and Information Engineering (DEI), University of Bologna, Italy}

\IEEEauthorblockA{\IEEEauthorrefmark{2}IIS Integrated Systems Laboratory, ETH Zurich, Switzerland}

\thanks{Manuscript received December 1, 2012; revised August 26, 2015. 
Corresponding author: A. Garofalo (email: angelo.garofalo@unibo.it).}}

\markboth{Journal of \LaTeX\ Class Files,~Vol.~14, No.~8, August~2015}%
{Shell \MakeLowercase{\textit{et al.}}: Bare Demo of IEEEtran.cls for IEEE Transactions on Magnetics Journals}
%



\IEEEtitleabstractindextext{%
\begin{abstract}
Strongly quantized fixed-point arithmetic is now considered a well-established solution to deploy Convolutional Neural Networks (CNNs) on limited-memory low-power IoT end-nodes.
Such a trend is challenging due to the lack of support for low bitwidth fixed-point instructions in the Instruction Set Architecture (ISA) of state-of-the-art embedded Microcontrollers (MCUs), which are mainly based on closed ISA such as ARM Thumb2 and associated Helium extensions. Emerging open-source ISAs such as RISC-V provide a flexible way to address this challenge.
This work introduces lightweight extensions to the RISC-V ISA to boost the efficiency of heavily Quantized Neural Network (QNN) inference on microcontroller-class cores. 
By extending the ISA with nibble (4-bit) and crumb (2-bit) SIMD instructions, we are able to show near-linear speedup with respect to higher precision integer computation on the key kernels for QNN computation.
Also, we propose a custom execution paradigm for SIMD sum-of-dot-product operations, which consists of fusing a dot product with a load operation, with an up to 1.64 $\times$ peak MAC/cycle improvement compared to a standard execution scenario.
To further push the efficiency, we integrate the RISC-V extended core in a parallel cluster of 8 processors, with near-linear improvement with respect to a single core architecture. To evaluate the proposed extensions, we fully implement the cluster of processors in GF22FDX technology. 
QNN convolution kernels on a parallel cluster implementing the proposed extension run 6 $\times$ and 8 $\times$ faster when considering 4- and 2-bit data operands, respectively, compared to a baseline processing cluster only supporting 8-bit SIMD instructions. With a peak of 2.22 TOPs/s/W, the proposed solution achieves efficiency levels comparable with dedicated DNN inference accelerators, and up to three orders of magnitude better than state-of-the-art ARM Cortex-M based microcontroller systems such as the low-end STM32L4 MCU and the high-end STM32H7 MCU.
\end{abstract}

}

\maketitle

\IEEEdisplaynontitleabstractindextext

%
\IEEEpeerreviewmaketitle

\section{Introduction}

In the last years we are assisting at an exponential growth of the Internet-of-Things (IoT) interconnected devices pervading several application domains such as agriculture, health monitoring \cite{hassanalieragh2015health}, structural health monitoring \cite{tokognon2017structural}.
This scenario requires the IoT end-nodes to acquire data from low-power sensors and send it wirelessly through the network\cite{li20185g}, after applying signal processing algorithms.

Machine Learning (ML) algorithms, including state-of-the-art Deep Learning (DL), not only empower the IoT nodes with smart capabilities widening the IoT applications with DL-enhanced tasks (e.g., autonomous nanodrones \cite{palossi201964}) but they provide ``information distillation'' solutions to extrapolate actionable information from the raw data acquired by sensors.
Their capability of ``squeezing'' raw data in a much more semantically dense format (e.g., extracting classes, high-level features, symbols) allows the wireless transmission of a limited amount of condensed information, alleviating the traffic on the IoT network and reducing the security and reliability issues, nowadays exacerbated by the significant increase of raw data flowing through the network \cite{shi2016edge}.
%

The clear benefits of embedding intelligence on IoT nodes have attracted the attention of a wide research area, with the goal of deploying DL functionality at the extreme-edge of the IoT.
This effort has to run against the high computational and memory requirements of leading DL methods that clash with the usual scarcity of computing and memory resources of deeply embedded systems powered by batteries or energy harvesters.

Typically, deep neural network (DNN) tasks run on GPUs or FPGAs devices in data centers, characterized by a power envelope that is orders of magnitude higher than what is sustainable on extreme-edge devices.
Dedicated DNN accelerators are starting to gain traction for ultra-low power devices  \cite{reuther2019survey,  daly2020through}, but they require extra silicon area, which is often not affordable in the extremely cost-conscious and fragmented IoT market. 
On the other hand, resource-constrained MCUs are the standard computing platforms used to build extreme-edge nodes, thanks to their flexible software programmability, low-cost and low-power characteristics.
However, they present severe limitations in memory footprint and computing resources, and these limitations may prevent meeting latency and accuracy requirements of the target DL-enhanced applications.

To reduce the model size of modern DNN topologies and make them fit the limited storage capacity of MCU-class devices, recent progress in DL training methodologies has introduced novel quantization methods.
These techniques represent the network weights and activations with 8-bits (or even smaller) data types, usually adopting fixed-point formats, incurring a limited or negligible loss in accuracy\cite{hubara2017quantized,lin2016fixed,wang2019haq,jacob2018quantization}.
%
The authors of \cite{rusci2019memory}, for example, show that the weights and the activations of a MobilenetV1 can be efficiently quantized to 8- or 4-bits with a loss on Top1 accuracy of only $0.8\%$ and $3.2\%$, respectively, compared to the fully fixed-point precision. At the same time, this approach reduces the memory footprint by $4 \times$ (8-bits) and by $7 \times$ (4-bits).

The limited footprint and the good accuracy achieved make Quantized Neural Networks (QNNs) the natural target to embed intelligence on MCU-based platforms and encourage many efforts by industry and academia to enable the QNN computation on microcontroller-based systems.
%
In this context, it is worth citing the CMSIS-NN library \cite{lai2018cmsis} developed by ARM for 16-bit and 8-bit QNNs on Cortex-M based systems and PULP-NN \cite{garofalo2020pulp}, an open-source library \footnote{https://github.com/pulp-platform/pulp-nn.git} targeting RISC-V processors and supporting heavily quantized NNs on 8-, 4-, 2-, 1-bits data, as well as Mixed-Precision QNNs \cite{bruschi2020enabling, garofalo2019pulp}. 

On the hardware side, modern MCUs show an inherent limitation that reduces the computing efficiency of such systems on sub-byte QNNs. 
This is due to the lack of support at the Instruction Set Architecture (ISA) level for low-bitwidth integer Single-Instruction-Multiple-Data (SIMD) arithmetic instructions; modern MCUs adopting commercial ISAs only support 16-bits (e.g., ARMv7E-M) or 8-bits (e.g., RV32IMCXpulpV2 \cite{gautschi2017near}, ARMv8.1-M \cite{M55}) data.

Hence, sub-byte quantization is an effective technique to compress the footprint of DNN models on top of these devices  \cite{rusci2019memory} but does not allow to reduce latency and energy consumption of the computation, as demonstrated in \cite{garofalo2020pulp, rusci2018work, capotondi2020cmix}. On the contrary, the lack of hardware support introduces a non-negligible overhead since low precision data have to be unpacked to the lowest precision operand supported by the underlying hardware and then packed into SIMD registers before feeding the multiply-accumulate (MAC) units \cite{bruschi2020enabling}.
In this work, we tackle this problem by proposing a set of lightweight domain-specific extensions to the RISC-V ISA, namely \textit{XpulpNN}, targeting specifically the computing requirements of low-bitwidth QNNs, with the support for sub-byte SIMD operations (8-, 4-, 2-bits).
%

The main contributions of this paper are the following:
\begin{itemize}
    \item We extend the RISC-V ISA with domain-specific instructions, namely \textit{XpulpNN}, for low-bitwidth integer SIMD computation, aiming at boosting the performance and the efficiency of heavily quantized QNN kernels; furthermore, we provide \textit{XpulpNN} with a class of mac\&load instructions, aiming at increasing the utilization of the SIMD MAC unit in the core towards the theoretical bound of 1 (0.92 in the best case scenario);
    \item We integrate the \textit{XpulpNN} extensions in the hardware design and on the register transfer level (RTL) description and the GCC toolchain of an open-source RISC-V processor \cite{gautschi2017near} already featuring specific extensions to target energy-efficient digital signal processing;
    \item We integrate the extended core in an eight cores parallel ultra-low-power (PULP) computing cluster, showing that we improve the performance of QNN kernels almost linearly with respect to the single-core execution; 
    \item We implement the PULP cluster integrating the proposed core in GF 22nm FDX technology to evaluate the area, power, and performance overhead of the core and the whole system with respect to the baseline RI5CY core and the PULP cluster integrating it, respectively; 
    \item We compare a PULP system with the proposed extension with state-of-the-art architectures and software. When running QNN convolution layers, we are able to demonstrate at least two orders of magnitude better performance and energy efficiency with respect to commercially available solutions such as STM32H7 and STM32L4 microcontrollers leveraging ARMv7E-M ISA, and up to 10$\times$ better performance and energy efficiency compared to a baseline PULP system implemented in the same technology, paying an area overhead of only 17.5\% adn 4.1\% with respect to the baseline core and cluster respectively.
    \end{itemize}

The orders of magnitude improvements achieved with the proposed work compared with state-of-the-art MCUs demonstrate for the first time that software programmable edge inference of QNN models at ASIC-like efficiency is indeed possible on MCU-class devices, with an additional cost of area which is risible if compared to the one of dedicated accelerators. These results can be achieved by coupling architectural and power-aware micro-architectural design with leading-edge near-threshold FDX technology.

\section{Related Work}
\label{sec:related_work}
The demand from the Internet-of-Things (IoT) for more embedded intelligence has paved the way for many different DL deployments on embedded computing platforms of all kinds. The edge Artificial Intelligence (AI) computing platforms can be grouped into three main categories: dedicated accelerators, FPGA solutions, and embedded Microcontroller (MCU) systems. In this section, we recap the state-of-the-art solutions and give insight on their applicability for DL deployment at the extreme-edge of the IoT. In this review, we leave out GP-GPUs, certainly valuable for Cloud Computing environment, but with a power envelope unaffordable for deeply embedded edge computing platforms working in a power envelope ranging from 10 mW to 100 mW.


\subsection{Dedicated Accelerators}
Dedicated accelerators are top-in-class for what concerns performance and energy efficiency on the QNN workloads. 
Having a highly specialized datapath, they can achieve performance in the order of 1 - 10 Gops/s with efficiency in the range of 10 - 100 Tops/s/W. Origami \cite{cavigelli2016origami}, for example, reaches a peak performance of 274 Gops/s with an energy efficiency of 803 Gops/s/W. An other valuable example is Orlando \cite{desoli201714}, which reaches few TOPS/W of efficiency.

The arithmetic precision drop is a valuable technique to further improve the QNN efficiency also on ASIC accelerators \cite{moons2017minimum}.
%
UNPU \cite{lee2018unpu} is an example of an accelerator supporting fully-variable weight bit-precision and capable of achieving a peak energy efficiency of 50.6 TOPS/W at a throughput of 184 GOPS. YodaNN \cite{andri2017yodann} targets binary-weighted networks with higher-precision activations reaching 61 TOPS/W of efficiency.
Moons et al. \cite{moons201714} presented ENVISION, an energy-scalable multi-precision DNN accelerator delivering 76 Gops/s with an efficiency up to 10 Tops/s/W.

The high performance and energy efficiency achieved by these accelerators are counterbalanced by their poor flexibility, which makes the end-to-end deployment of real-sized DNNs harder.
Moreover, even if modern dedicated architectures have a datapath somehow reconfigurable (for example, allowing the execution of convolutions with different kernel sizes, 3x3, 5x5 or they provide the possibility to handle inception layers and/or residual connections), they can not be configured to support different kind of applications.
In the IoT domain, instead, this flexibility is crucial. The DNN inference is usually only one part of a bigger application, where we additionally may want to handle peripherals, process the data through linear algebra, domain-to-domain transforms (even recurring to floating-point numbers) and manage the wireless transmission of the high-level compressed results. The poor flexibility and the high-cost per device make the ASIC solutions unattractive for their use as sensor-nodes at the extreme edge of the IoT.
%
%

\subsection{FPGAs}
The recent development of heterogeneous FPGAs such as the Xilinx Zynq family [] has enabled a higher level of flexibility to build CNN acceleration systems. Embedding general purpose processors on the FPGA boards allow managing the program flow, handling the I/O sub-system, memory accesses, communication, hence making easier to program the device and interact with external devices and sensors.

FPGAs usually come with DSP-capable hardware, but they have a power envelope in the order of the Watt. Thus the reduction of numerical precision for CNN models plays a key role in achieving good performance and energy efficiency.
In the literature, we can find several FPGA-based solutions that exploit 16-bit fixed-point operands such as in \cite{gokhale2017snowflake, ma2017automatic, meloni2018neuraghe}, but an ever increasing number of works explore byte or sub-byte arithmetic. Qiu e al. \cite{qiu2016going} proposed a CNN accelerator supporting 8- and 4-bit data on a Xilinx Zynq board, while \cite{prost2017scalable, umuroglu2017finn} rely on ternary and binary networks.
While most FPGA solutions feature a power envelope that can not meet the IoT end-nodes requirements, a new family of FPGAs announced by Lattice, namely SenseAI \cite{LatticeSENSEAI}, provide comprehensive hardware and software solutions for always-on artificial intelligence (AI) within a power budget between 1 mW and 1 W. 

However, these ultra-low-power FPGA families feature limited LUTs capability, and still are too expensive for many applications where MCUs are traditionally chosen thanks to their low cost. On the other hand, their efficiency remains way lower than what ASICs can offer.
In addition, they can be reconfigured using a Hardware Description Language (HDL), increasing the productivity with respect to the above mentioned ASIC solutions; still, their adoption remains an obstacle for the average IoT programmer, who demands for the highest flexibility of microcontroller systems.

\subsection{Software Programmable solutions}

Commercially available software-programmable general-purpose processors provide the highest flexibility for the deployment of the QNN at the extreme-edge. However, the ultra-low-power MCUs are not fast enough, if compared to ASICs, to sustain the QNN workload.
Some programmable architectures, for example, exploit the computing power of multi-core processors, such as Raspberry Pi 3+ \cite{Raspberry} or Pi 4, powered by a multi-core 64-bit System-on-Chip by ARM. 
Other solutions couple a programmable core with a dedicated accelerator to improve performance, such as Kendryte \cite{Kendryte}, a dual-core RISC-V SoC outfitted with a CNN accelerator for AI applications, or more established devices like Movidius \cite{movidius} or the Edge TPU \footnote{https://cloud.google.com/edge-tpu} coupled with a CPU.
Although these platforms are relatively inexpensive and flexible, their power consumption is too high as well.
In the ultra-low-power domain instead, ARM proposed Trilium \cite{trilium}, a heterogeneous compute platform which provides flexible support for ML workloads. Conti et al. \cite{conti2015ultra} proposed a convolution engine to be integrated in a microcontroller to speed up the convolutional kernels. Also in this case, the hybrid solutions provide flexibility and efficiency, but the increased cost (due to the dedicated accelerator hosted) in terms of area is not negligible. In many IoT applications the cost constraints are very tight  and it is desirable to reduce area while having an MCU with boosted computing capabilities.

To enhance the performance of low power MCU systems, a recent effort by both academia and industry tries to extend them by either enriching their Instruction Set Architectures (ISAs) with custom instructions tailored for specific application domains or coupling the MCUs with ASIC accelerators.

ARMv7e provides SIMD instructions for 16-bit data, and the current generation of Cortex-M cores integrate this instruction set. Commercial embodiments of this ISA show power envelope of few milliWatts, fitting the power budget of the IoT endnodes. For example, STMicroelectronics proposed low-end (STM32L4 \footnote{ https://www.st.com/resource/en/datasheet/stm32l476je.pdf} family of microcontrollers, based on the Cortex-M4 cores) and high-end (STM32H7 \footnote{https://www.st.com/resource/en/datasheet/stm32h743bi.pdf} family embedding the Cortex-M7 cores) microcontrollers supporting DL processing at the edge.
On the RISC-V side, the XpulpV2 ISA extensions \cite{gautschi2017near} are meant for efficient digital signal processing, exploiting the SIMD paradigm down to 8-bit vector data. Andri et. al. propose \cite{andri2020extending} efficient RISC-V extensions for Recurrent Neural Network (RNN) computation. On top of this ISA, near-threshold multi-core heterogeneous platforms have been built to push the performance and the efficiency of QNN workloads \cite{pullini2019mr}. The commercially available GAP-8 \cite{flamand2018gap} embeds a cluster of 8 RISC-V cores and a CNN-specialized accelerator that can give the MCU a 5 to 10$\times$ energy efficiency boost.

To provide these architectures with an efficient software back-end, several solutions are presented in literature that achieves promising results on QNN workloads. It is worth citing the CMSIS-NN \cite{lai2018cmsis} library, developed by ARM to target their Cortex-M cores. As an additional contribution, this library has been extended to support heavily-quantized and mixed-precision kernels \cite{rusci2018work, capotondi2020cmix}.
On the RISC-V side, PULP-NN \cite{garofalo2020pulp, bruschi2020enabling} provides a solid back-end for RISC-V based multi-processors systems, supporting byte as well as sub-byte and mixed-precision QNN kernels.

Usually, these libraries are integrated into a vertical deployment flow, which also takes care of memory management during computation.
X-CUBE-AI \footnote{} from STMicroelectronics is an automatic library generator for Neural Networks optimized on computation and memory. It takes as input a DNN model from most used commercial tools (Keras, Tensorflow) and generates a C-like library ready to be executed on top of the ARM Cortex-M cores, embedded in STM32 devices, using the CMSIS-NN library.
Dory \cite{burrello2020dory} instead is a general memory oriented deployment framework for DNNs that are deployable on any edge system (e.g., ASICs and MCUs) by specifying some heuristics. It takes as input an ONNX graph of the network and generates the binary to be run on the system. Specialized for a Parallel-Ultra-Low-Power (PULP) system such as GAP-8 and embedding the PULP-NN primitives as backend, Dory demonstrates an efficient end-to-end inference of an 8-bit Quantized Mobilenet V1.

Even if the sub-byte integer arithmetic is already adopted in training and quantization flows and ASIC/FPGA -based systems, it does not find enough room in the new generation of architectural solutions for MCU-based systems.
From a hardware perspective, a relevant work is the reconfigurable Parallel Balanced-BitSerial (PBBS) vector processing tile \cite{wu2017parallel}. It is suitable to improve the efficiency of sub-byte SIMD computation of heavily leakage-dominated ultra-low-power design. However, the code serialization degrades heavily the performance in near- and super-threshold operating points. Moreover, commercial MCUs operate at the finest granularity of the byte.
The new generation of the ARM ISA for Cortex-M core \cite{M55}, tailored for the QNN workload, features hardware loops, conditional execution instructions, and 8-bit SIMD instructions like the ones presented in \cite{gautschi2017near}, but it will not support lower-precision SIMD arithmetic.

Our work aims at bridging this ISA and hardware gap to improve the computing efficiency of heavily-quantized NN workloads at the extreme-edge of the IoT on fully-programmable MCU devices, nearing the level of specialization and energy efficiency of custom accelerators without forgoing flexibility.
To this purpose, we extend the work presented in \cite{garofalo2020xpulpnn} and propose a set of RISC-V ISA extensions, consisting of low-bitwidth SIMD arithmetic operations and a family of \emph{mac\&load} instructions, coupled with a parallel ultra-low-power (PULP) architecture to achieve high energy efficiency. 
By exploiting the proposed ISA extensions and their integration in a tightly coupled cluster of 8 processors, our contribution outperforms the state-of-the-art hardware and software solutions by at least two orders of magnitude in terms of performance and efficiency.

To  put  our  results  into  perspective,  Table  I  provides  a  sum-mary  of  embedded  computing  platforms  for  QNNs  targetingan  operating  power  below1 Wwith  our  proposed  work.

\begin{table}[t]
\caption{Overview of QNN Embedded Computing Platforms and Main Metrics}
 \label{tab:related_work}
  \begin{tabular}{ccccc}
     \hline
                                                    & Throughput  & Energy     & Power   & Flexibility \\
                                                    &             & Efficiency & Budget  &             \\
                                                    & [Gop/s]    & [Gop/s/W] & [mW]    &             \\
     \hline 
     \textbf{ASICs \cite{desoli201714,lee2018unpu}} & 1K - 50K    & 10K - 100K & 1 - 1K  & Low         \\
     \hline
     \textbf{FPGAs \cite{LatticeSENSEAI}}           & 10 - 200    & 1 - 10     & 1 - 1K  & Medium      \\
     \hline
     \textbf{MCUs \cite{garofalo2020pulp, garofalo2020xpulpnn}}        & 1 - 5     & 80 - 550     & 1 - 1K  & High        \\
     \hline
     \textbf{This Work}                             & 40 - 140       & 1k - 3k   & 1 - 100 & High        \\
    \hline

\end{tabular}
\end{table}

\section{Background}
\subsection{Quantized Neural Networks (QNNs)}
\label{sec:qnns}
Quantized Neural Networks (QNNs) are the result of post-training quantization~\cite{capotondi2020cmix} or quantization-aware training~\cite{choi2018pact} procedures.
After the quantization, each tensor $\mathbf{t}$ of the QNN (e.g., weights $\mathbf{w}$, input activations $\mathbf{x}$, or outputs $\mathbf{y}$) can assume only a finite set of values which are defined in a specific real-valued range $[\alpha_\mathbf{t}, \beta_\mathbf{t})$.
These discretized real values can be mapped, through bijective functions, into pure integer numbers called \textit{integer images} of the real-valued discretized tensors.
More in detail the $N$-bit integer image $\widehat{\mathbf{t}}$ of the tensor $\mathbf{t}$ is connected to its real-valued quantized counterpart through the following function:
\begin{equation}
    \mathbf{t} = \alpha_\mathbf{t} + \varepsilon_\mathbf{t}\cdot \widehat{\mathbf{t}} \label{eq:1}\;, 
\end{equation}
where $\varepsilon_\mathbf{t} = (\beta_\mathbf{t}-\alpha_\mathbf{t}) / (2^{N}-1)$.
We call $\varepsilon_\mathbf{t}$ the \textit{quantum} because it is the smallest amount that we can represent in the quantized tensor. Without loss of generality, we further constraint $\alpha_\mathbf{x} = \alpha_\mathbf{y} = 0$  for the input activations and the output features of each QNN layer.
After mapping all the tensors in the integer domain, the application of the QNN operators (Linear Operator, Batch-Normalization, and the Quantization/Activation) can operate directly on the \textit{integer images}:

    \begin{align}
        \mathrm{LIN:} \quad&\varphi = \sum_n \mathbf{w}_{m,n} \mathbf{x}_n \iff \widehat{\varphi} = \sum_n \widehat{\mathbf{w}_{m,n}} \cdot \widehat{\mathbf{x}_n}
        \label{eq:2} \\
        \mathrm{BN:} \quad&{\varphi'} = {\kappa}\cdot {\varphi}+{\lambda} \iff \widehat{\varphi}' = \widehat{\kappa}\cdot \widehat{\varphi}+\widehat{\lambda}
    \label{eq:normalization}
        \;.
    \end{align}

In the LIN operator, the accumulator of the dot product operation will be represented, in general, with higher precision (e.g., 32 bits) with respect to the two inputs, since the quantum used to represent the accumulator $\widehat{\varphi}$ will be smaller than that of the two operands ($\varepsilon_\varphi = \varepsilon_\mathbf{w}\varepsilon_\mathbf{x}$).
The same consideration also holds for the output of the Batch-Normalization operator. The final Quantization/Activation operator provides a non-linear activation semantic, which is essential for QNNs to work, and collapses the accumulator into a smaller desired bitwidth:

\begin{equation}
    \mathrm{QNT/ACT:} \quad\widehat{\mathbf{y}} = 
    m \cdot \widehat{\varphi}' \gg d \;;\;
    m=\left\lfloor\frac{
        \varepsilon_{\mathbf{\varphi}'}\cdot 2^{d}
    }{
        \varepsilon_\mathbf{y}
    }\right\rfloor\;. 
    \label{eq:requantization}
\end{equation}

$d$ is an integer chosen during the quantization process in such a way that $\varepsilon_\varphi/\varepsilon_\mathbf{y}$ can be represented with sufficient accuracy inside $m$.
The BN and QNT/ACT operators can also be implemented through a stair-case function by folding the BN and QNT/ACT parameters into a set of thresholds.
The staircase-function compares $\varphi$ with a set of $2^N$ thresholds to compress the result into N bits, with a computational complexity of $O(N)$.
To implement the quantization with a thresholding-based method, we would need to store $2^N$ thresholds per output channel, which leads to a large memory footprint for real-world convolution kernels.
Since the computational complexity is comparable between the two methods for real-world layers, we will always assume in the rest of the manuscript that the Quantization and Normalization steps are implemented with the BN and QNT/ACT operators, as explained in this section.

\subsection{PULP cluster}
\label{sec:PULP_RISCY}
\begin{figure}
    \centering
    \includegraphics[width=0.97\linewidth]{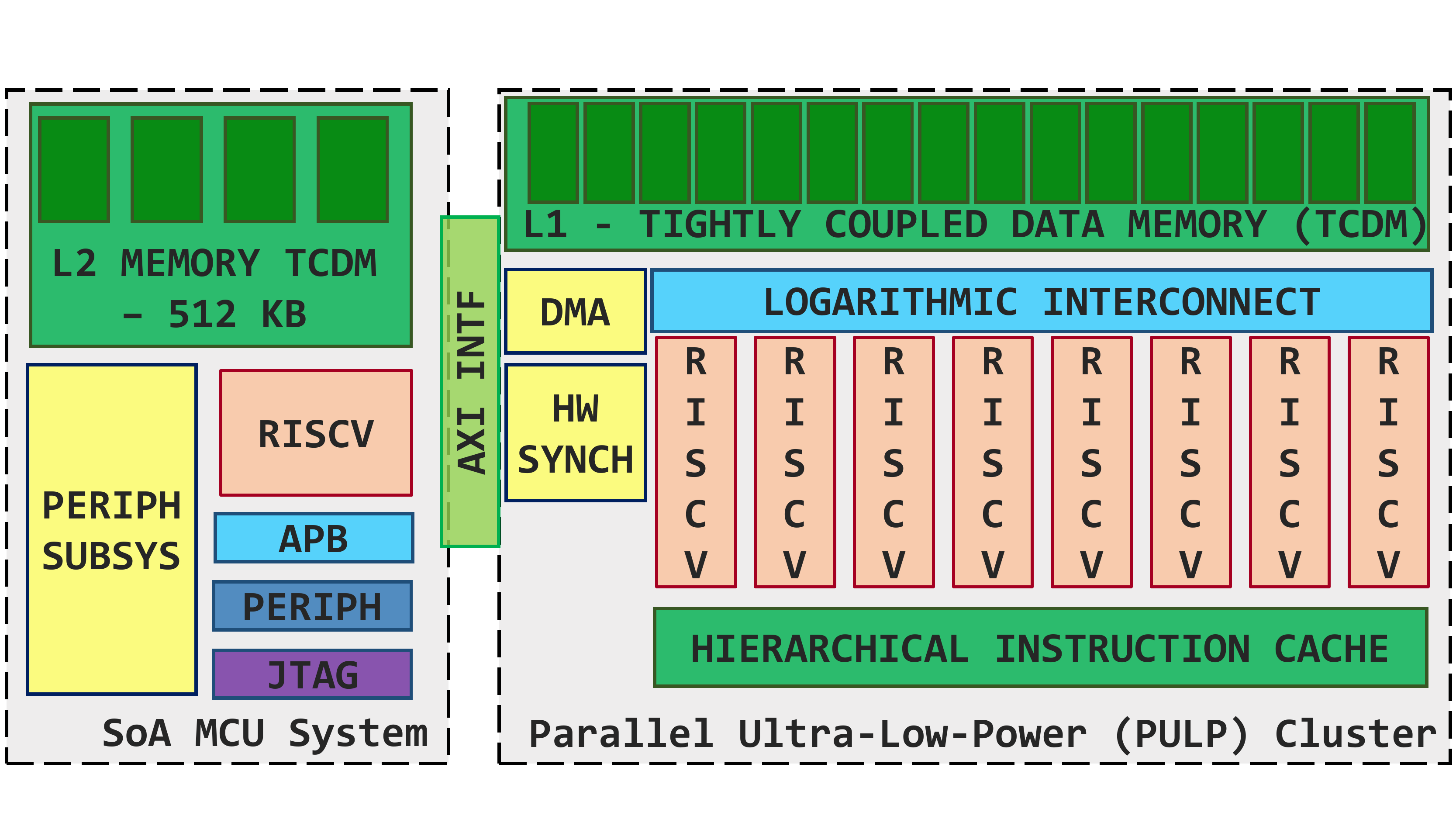}
    \caption{The Parallel Ultra-Low-Power (PULP) system, consisting of a state-of-the-art microcontroller system accelerated by a parallel cluster of 8 RISC-V based processors.}
    \label{fig:pulp_cluster}
\end{figure}
Parallel Ultra-Low Power (PULP) is an open-source computing platform leveraging near-threshold computing to achieve
high energy efficiency, leveraging parallelism to improve the performance degradation at low-voltage \cite{pullini2019mr}.
The PULP cluster we assume as a reference, depicted in Figure \ref{fig:pulp_cluster}, is composed of eight RI5CY cores \cite{gautschi2017near}, each featuring a 4-stage in-order single-issue pipeline and implementing the RISC-V RV32IMCXpulpV2 Instruction Set Architecture (ISA). 

The XpulpV2 is a custom extension to the RISC-V ISA \cite{gautschi2017near} meant for efficient digital signal processing computation. To this purpose, it includes hardware loops, post-modified access load and store instructions, as well as the support for SIMD operations down to 8-bit integer vector operands.

The cores of the baseline cluster synchronize through a shared Tightly Coupled Data Memory (TCDM) of 128 kB, divided on multiple-banks with a banking factor of two (i.e., 16 banks for the 8-cores configuration). The cores access the memory through a low latency logarithmic interconnect that serves the memory accesses in one cycle. 

Meant to accelerate a microcontroller system, the PULP cluster communicates with its host through an AXI interface. It is also served with a DMA which dedicated to the data transfers between the TCDM and the second level of memory, hosted by the microcontroller system, which also contains the program instructions for the cluster cores. 

Each core fetches the instructions from a hierarchical instruction cache organized on two levels (the first private to each core, the second shared) to optimize the hit rate. The cluster is also provided with a Hardware Synchronization Unit that manages synchronization and thread dispatching, enabling low-overhead and fine-grained parallelism, thus high energy efficiency: each core waiting for a barrier is brought into a fully clock gated state.

\subsection{QNN Execution Model}
\label{sec:QNN_EXEC_MODEL}

The software stack used in this paper to assess the results of the ISA extensions and the architectural explorations is derived from the open-source PULP-NN library\footnote{https://github.com/pulp-platform/pulp-nn}, tailored for optimized parallel execution of QNN kernels on PULP and the above mentioned XpulpV2 ISA.
In this work, we further extend the sub-byte symmetric convolution kernels of this library with the \textit{Xpulpnn} ISA instructions.

The PULP-NN library relies on the Height-Width-Channel (HWC) data layout and on an execution flow optimized for resource-constrained microcontrollers. A convolution layer is implemented as a combination of three distinct phases.
\subsubsection*{\textbf{im2col}} This step takes the 3-D input activations in format (H,W,C) and, for a given output position, arranges its full receptive field along the filter and input channel dimensions into a 1-D vector. In this way, the full convolutional layer operation is converted into a scalar product between this vector and flattened weights. PULP-NN performs this operation for 2 output pixels concurrently, creating two distinct \textit{im2col buffers} of $C_{in}\times F\times F$ elements each.
\begin{figure*}
  \centering
  \subfigure[\centering Layout and hardware resources (registers) of the ``4$\times$2'' and the ``4$\times$4'' layouts of the \textit{MatMul} kernels.]{\includegraphics[width=.45\linewidth]{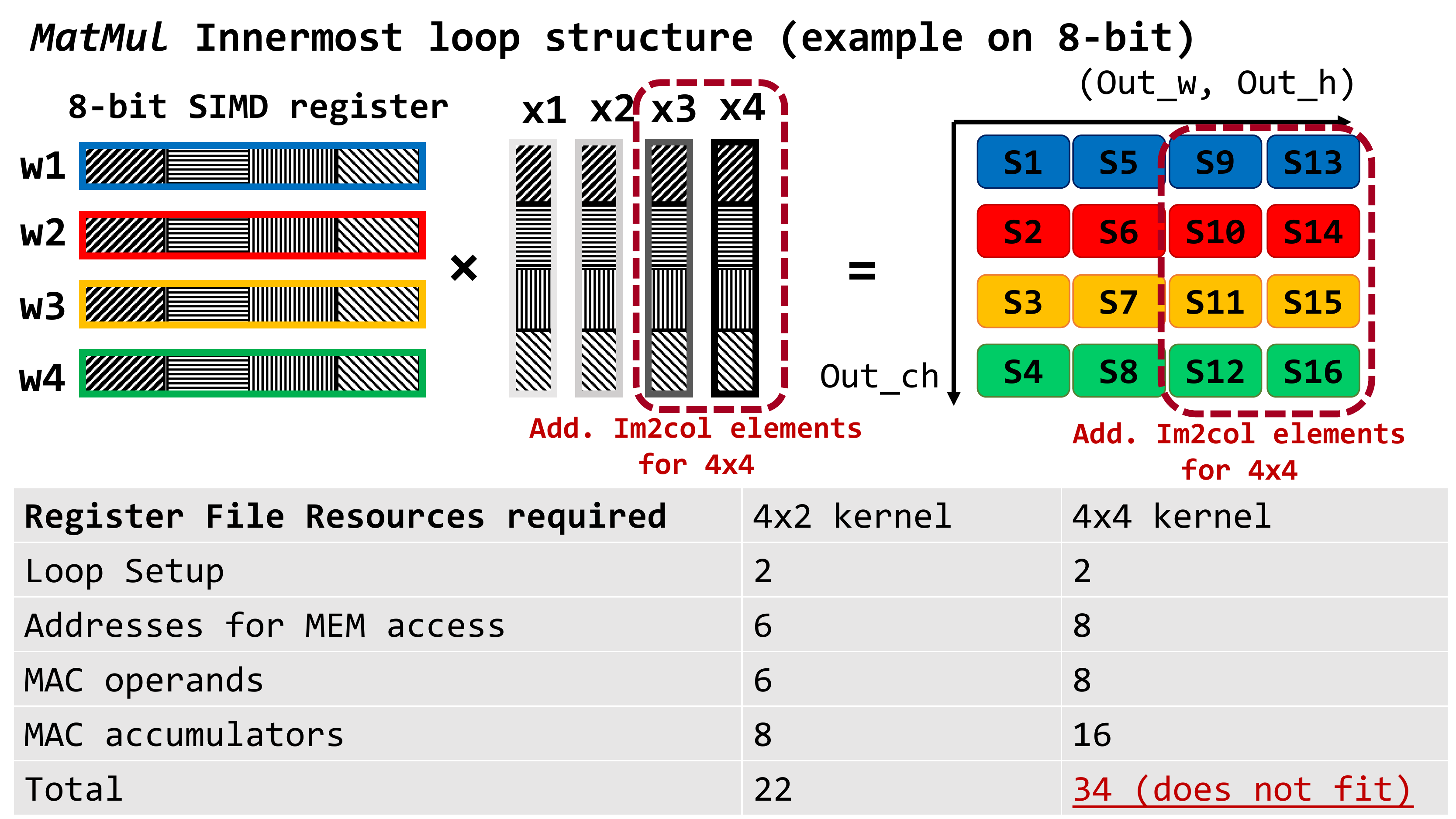} }%
    \qquad 
    \subfigure[\centering Assembly code of the innermost loop of the ``4$\times$2'' \textit{MatMul} kernel . The figure puts in perspective the RF resources needed to run the loop.]{\includegraphics[width=.45\linewidth]{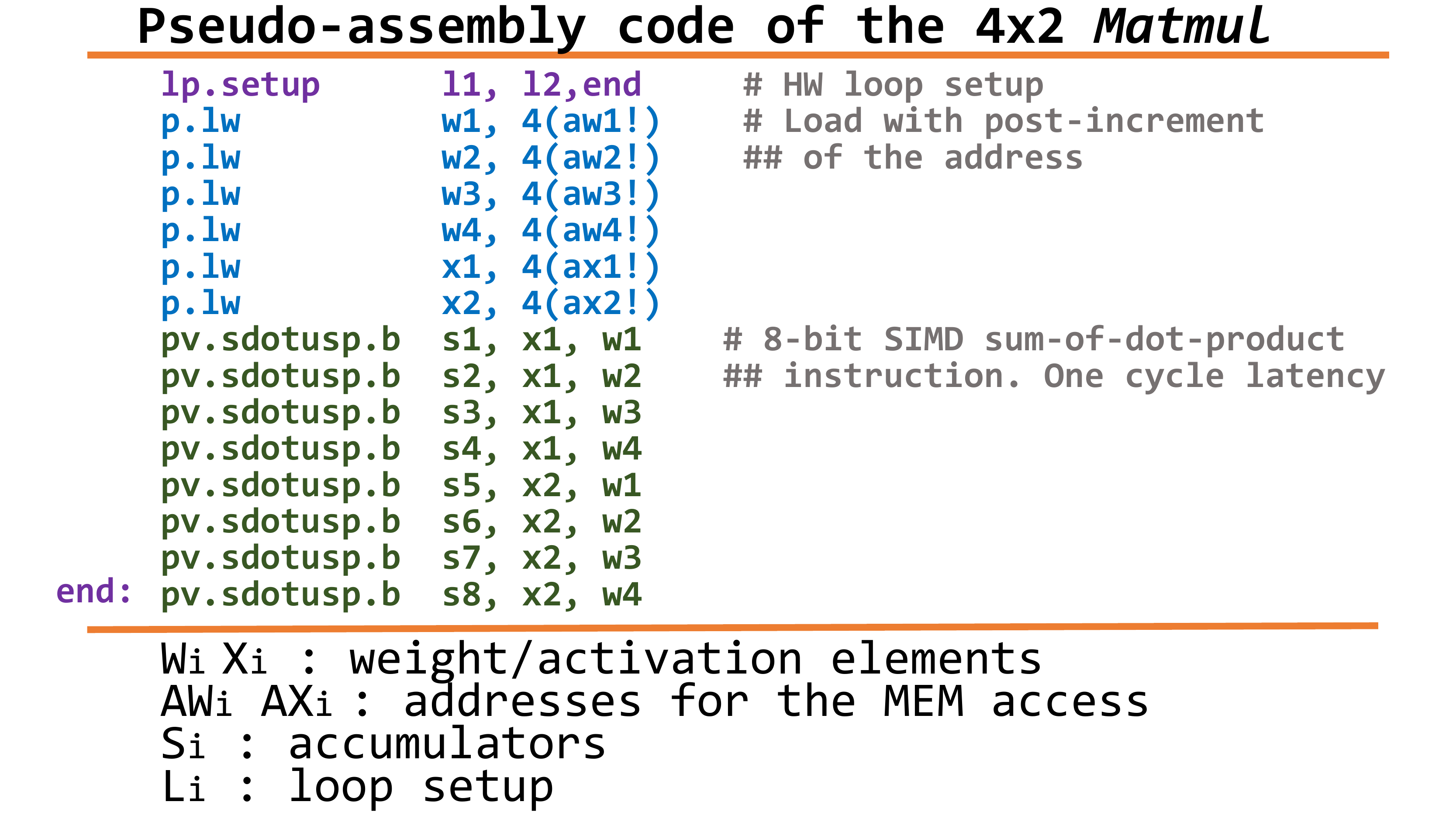} }%
  \caption{The Figure shows the  structure and the assembly code of innermost loop of the \textit{MatMul} kernel of the PULP-NN library. The ``4$\times$2'' kernel structure fetches two activations ($x_1$ and $x_2$) from two different \textit{im2col} buffers and the weights ($w_1$ to $w_4$) from four different filter sets to compute eight intermediate results ($s_1$ to $s_8$), requiring 22 registers available in the RF of RI5CY. The ``4$\times$4'' layout can not be implemented on RI5CY, since the registers needed for the computation would not fit efficienctly the RI5CY register file. }
  \label{fig:matmul_pulpnn}
\end{figure*}
\subsubsection*{\textbf{Matrix Multiplication}} This step is the core of convolution, and it performs a sum-of-dot-product operation between the current \textit{im2col buffer} and the sets of filters to produce higher precision results (the intermediate values of the output activations), which usually feature 32-bits. The kernel is highly optimized with the XpulpV2 instructions: hardware loops (\textit{lp.setup}), load with post-increment (\textit{p.lw}) and the SIMD \textit{sdotp} (sum-of-dot-product) instruction which delivers 4 MACs in one cycle latency on 8-bit SIMD operands.
As shown in Figure \ref{fig:matmul_pulpnn}, to optimize performance the \textit{MatMul} uses activations from two \textit{im2col} buffers (associated to two spatially adjacent output pixels) and the quantized weights from four filter banks associated to four output channels.
    Exploiting the data locality within the RF enables the computation of eight output pixels per each iteration of the \textit{MatMul} inner loop (4 channels $\times$ 2 adjacent pixels).
    This ``$4\times2$'' structure is the result of a design space exploration aiming at finding the data reuse condition maximizing throughput~\cite{garofalo2019pulp}.
    Due to the limited amount of slots available in the RI5CY register file to store operands and accumulators, no further reuse can be exploited on RI5CY -- in fact, wider \textit{MatMul} structures (e.g., 4$\times$4) would be detrimental as the additional accumulators would exceed the number of available registers, as visible from the Figure \ref{fig:matmul_pulpnn}.
    This causes the compiler to continuously spill operands back and forth from the stack, introducing significant overhead \cite{garofalo2019pulp};
\subsubsection*{\textbf{Quantization}} As discussed in Section~\ref{sec:qnns}, intermediate accumulators require 32-bit precision, and they need a final step of normalization and quantization to be represented in low-bitwidth form. These functions consist of one MAC operation, one shift, and one clip instruction per each accumulator to be quantized back into the desired precision.
    Contrarily to other quantization strategies such as thresholding-based quantization \cite{rusci2018work, rusci2019memory}, the computational complexity of using explicit integer Batch-Normalization does not depend on the output activation precision. As a result, the \textit{Quantization} stage adopted in this work will affect the overall performance to a higher degree when the precision is lower and the \textit{MatMul} is proportionally faster.
    Nevertheless, the advantage of this type of quantization resides in the lower memory footprint of its parameters with respect to thresholds (where per each output channel we would need to store $2^N$ thresholds for a final $N-$bit output activation). This behavior is more representative of the real-world QNN based tasks as shown, for example, in \cite{burrello2020dory}. 
    After normalization and quantization, the result is stored back into an 8-bit variable. For sub-byte operands, more output activations are compressed and stored back always into an 8-bit variable (which would contain either two 4-bit or four 2-bit elements) to reduce the memory footprint of the quantized output feature mapped.
\section{XpulpNN Extensions}

We present the SIMD instructions of the \textit{XpulpNN} ISA extensions and the micro-architecture design to support them in the baseline RI5CY core \cite{gautschi2017near}. Then, we introduce the concept of the Mac\&Load computation, presenting two different variants and comparing their benefits and their drawbacks. In the end, we integrate the RI5CY core extended with the new instructions into a parallel ultra-low-power cluster of eight processors, and we describe the software stack needed to execute the QNN convolution kernels on top of the \textit{XpulpNN} ISA.

\subsection{SIMD Extensions and Microarchitecture} 
\label{sec:ISA}
The RISC-V ISA we assume as our baseline, namely \textit{XpulpV2} \cite{gautschi2017near}, targets efficient digital signal processing exploiting SIMD computing paradigm on 16- and 8-bit vector operands.
The proposed \textit{XpulpNN} instructions, listed in Table~\ref{tab:xpulpnnisa}, extend the RV32IMCXpulpV2 ISA with SIMD operations for 4-bit and 2-bit operands, namely \textit{nibble} (indicated with \textit{n}) and \textit{\textit{crumb}} (indicated as \textit{c}) respectively, to improve the efficiency of low-bitwidth QNN kernels.
\begin{table}[t]
    \centering
        \caption{Overview of \textit{XpulpNN} instructions for \textit{nibble} (4-bit) and \textit{crumb} (2-bit) vector operands. $i$ in the table refers to the index in the vector operand ($i \in [0; 7]$ for \textit{nibble} and $i \in [0; 15]$ for \textit{crumb}). }
 
    \label{tab:xpulpnnisa}
    \begin{tabular}{l|l}
         \hline 
        \textbf{ALU SIMD Op.} & \textbf{Description for \textit{nibble}} \\
         pv.add[.sc].\{n, c\} &  rD[i] = rs1[i] + rs2[i]\\
          pv.sub[.sc].\{n, c\} & rD[i] = rs1[i] - rs2[i] \\ 
          pv.avg(u)[.sc].\{n, c\} & rD[i] = (rs1[i] + rs2[i])$>>$1\\
          \hline
          \textbf{Vector Comparison Op.}& \\
          pv.max(u)[.sc].\{n, c\} & rD[i] = rs1[i] $>$ rs2[i] ? rs1[i] : rs2[i]\\
          pv.min(u)[.sc].\{n, c\}& rD[i] = rs1[i] $<$ rs2[i] ? rs1[i] : rs2[i]\\ 
          \hline
          \textbf{Vector Shift Op.} & \\
          pv.srl[.sc].\{n, c\} &  rD[i] = rs1[i] $>>$ rs2[i] Shift is logical \\
          pv.sra[.sc].\{n, c\} & rD[i] = rs1[i] $>>$ rs2[i] Shift is arithmetic\\
          pv.sll[.sc].\{n, c\}& rD[i] = rs1[i] $<<$ rs2[i]\\ 
          \hline
          \textbf{Vector abs Op.} & \\
          pv.abs.\{n, c\}& rD[i] = rs1[i] $<$ 0 ? -rs1[i] : rs1[i]\\ 
          \hline
          \textbf{Dot Product Op.}& \\
          pv.dotup[.sc].\{n, c\}& rD = rs1[0]*rs2[0] + ... + rs1[7]*rs2[7]\\
          pv.dotusp[.sc].\{n, c\} & rD = rs1[0]*rs2[0] + ... + rs1[7]*rs2[7]\\
          pv.dotsp[.sc].\{n, c\} & rD = rs1[0]*rs2[0] + ... + rs1[7]*rs2[7]\\
          pv.sdotup[.sc].\{n, c\} & rD = rs1[0]*rs2[0] + ... + rs1[7]*rs2[7] + rD\\
          pv.sdotusp[.sc].\{n, c\}& rD = rs1[0]*rs2[0] + ... + rs1[7]*rs2[7] + rD\\
          pv.sdotsp[.sc].\{n, c\} &  rD = rs1[0]*rs2[0] + ... + rs1[7]*rs2[7] + rD\\
          \hline
    \end{tabular}
\end{table}

XpulpV2 supports three addressing variations: the first one uses two registers as source operands (\textit{pv.instr.\{b,h\}}), the second variation uses one  register and one immediate as source operands  (\textit{pv.instr.sci.\{b,h\}}), while the last one uses one register and replicates the scalar value in a register as the second operand for the SIMD operation (\textit{pv.instr.sc.\{b,h\}}).     
Because of the limited room left in the encoding space of the baseline ISA, we propose the new \textit{XpulpNN} crumb and nibble operations only in two addressing variants, and we do not implement the instruction format which uses an immediate value as the second operand (i.e., \textit{pv.instr.sci.\{b,h\}}). 
Based on our experience, we argue that this choice is not a concern for the execution of QNN kernels: an immediate value can be stored in advance into a register without additional overhead.

The core of the \textit{XpulpNN} ISA extension consists of the SIMD \textit{dot product} instructions on packed vectors of 4-, 2-bit elements. 
The packed input registers can be interpreted as both signed or unsigned, or the first signed and the second unsigned. The accumulator, as well as the third scalar input in the sum-of-dot-product, can be either signed or unsigned.
In addition to the \textit{dot product} we support other SIMD instructions like maximum, minimum, and average for nibble and crumb packed operands, useful to speed-up the pooling layers and the activation layers based on the Rectified Linear Unit (ReLu) function.
A group of arithmetic and logic operations (addition, subtraction, shift) completes the set of the \textit{XpulpNN} SIMD instructions.

%
\begin{figure}[t]
    \centering
    \includegraphics[width=\linewidth]{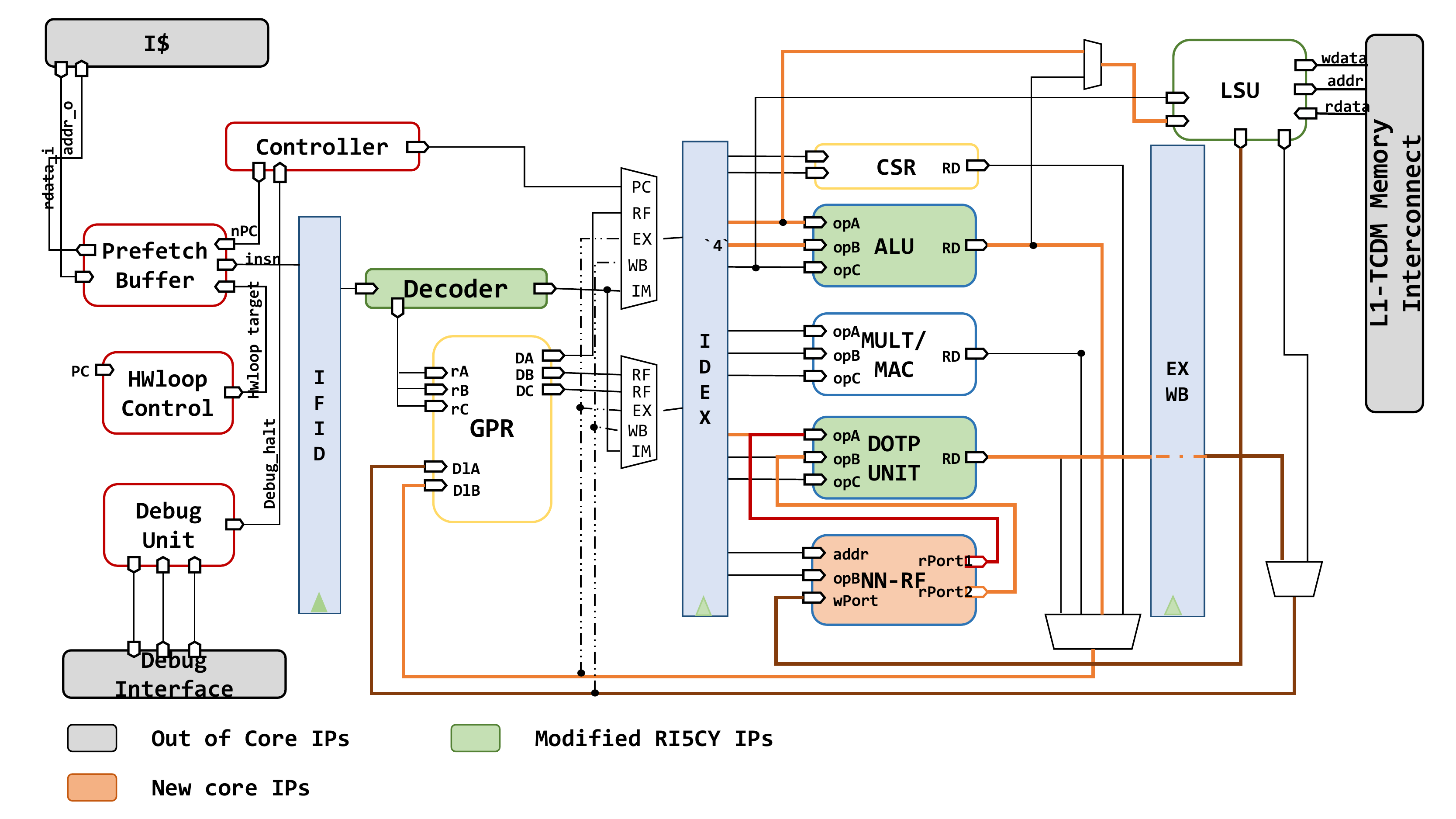}
    \caption{The RI5CY pipeline. The Figure highlights the hardware blocks which extend the core micro-architecture to support the \textit{XpulpNN} ISA.}
    \label{fig:riscy_pipelin}
\end{figure}

\begin{figure}[t]
    \centering
    \includegraphics[width=\linewidth]{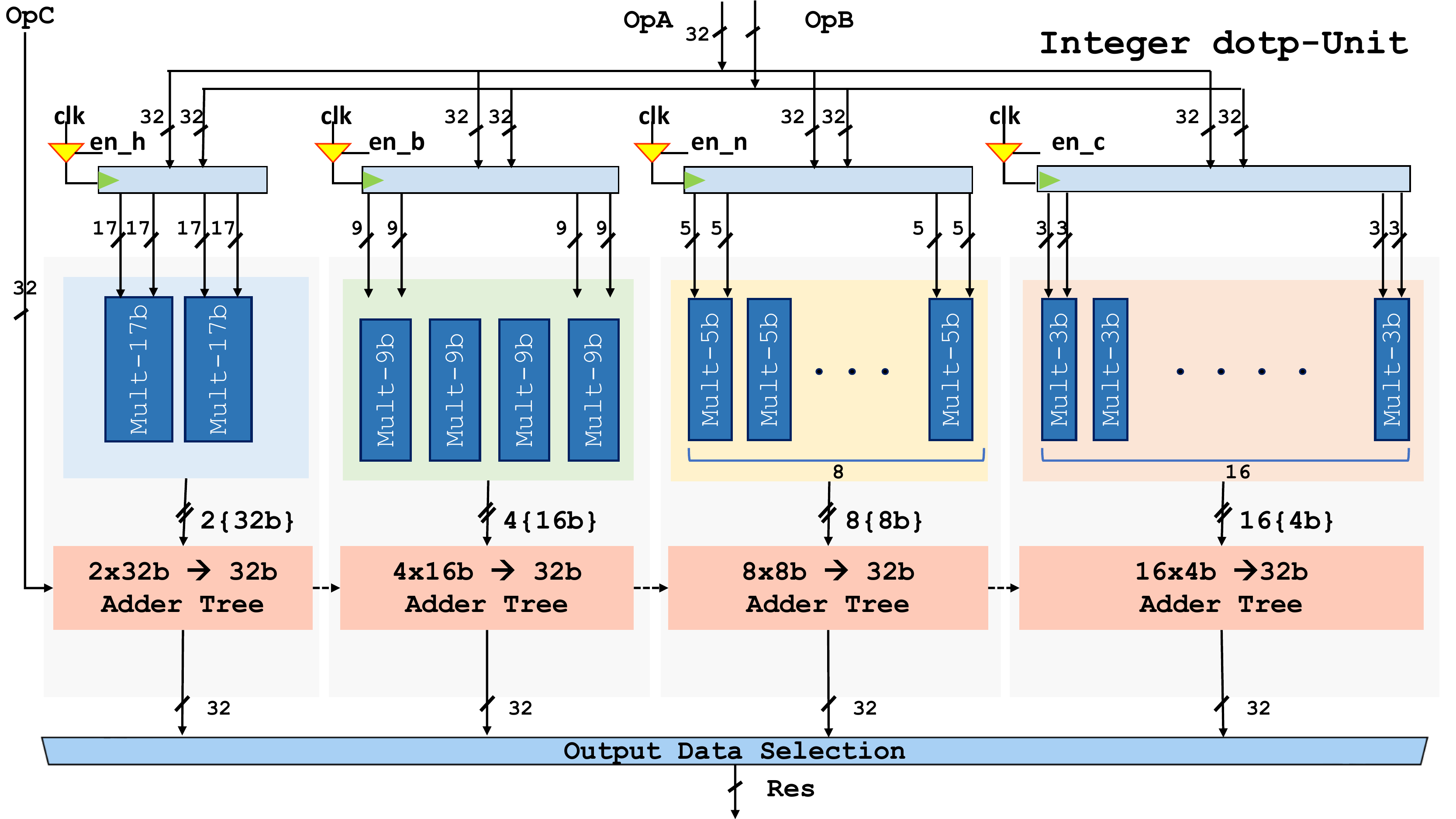}
    \caption{Block diagram of the RI5CY Dot-Product Unit. To support the \textit{XpulpNN} SIMD dotp-based operations, the 8$\times$4 and the 16$\times$2 SIMD MAC Units have been added. The figure includes the clock gating blocks needed to reduce the operand switching activity.}
    \label{fig:MAC_units}
\end{figure}

To support the SIMD instructions of \textit{XpulpNN}, we extend the micro-architecture of the baseline core (RI5CY \cite{gautschi2017near}) with the addition of additional multipliers into the Dot-Product unit and some logic in the ALU and in the decoder. The modified blocks are reported in green in Figure \ref{fig:riscy_pipelin}.
The baseline Dot-Product Unit available in the RI5CY core consists of two different sets of multipliers, one for 8-bit packed operands, the other for the 16-bit ones.
In the extended core, we support the \textit{XpulpNN} \textit{dotp} instructions by extending the Dot-Product unit with two additional sets of multipliers for nibble (4-bit) and crumb (2-bit) SIMD operands, as visible from Figure \ref{fig:MAC_units}.
The new sets of multipliers compute the dot product  operations between two SIMD registers, each containing either eight 4-bit or sixteen 2-bit elements, and accumulate the partial results over a 32-bit scalar register through an adder tree in one clock cycle of latency. 
The sum-of-dot-product (\textit{sdotp}) operation, which is the SIMD equivalent of a MAC operation, is supported by adding to the multipliers an additional 32-bit scalar operand at the input of each adder tree.
As reported in Section \ref{sec:ISA}, we support the nibble and crumb SIMD instructions interpreting the operands as signed or unsigned. Hence, we provide the inputs of the SIMD multipliers with an extra bit that sign- or zero-extends the actual single element of the SIMD vector. Therefore, as depicted in Figure \ref{fig:MAC_units}, each element is a 5- or 3-bit operand for the nibble and crumb cases, respectively.

Our design choice to replicate the hardware resources to support the \textit{XpulpNN} dot-product instructions aims at minimizing the impact of the additional hardware on the critical path of the  RI5CY core, which already involves the Dotp-Unit.
The nibble and crumb \textit{dotp} operations are near to be timing critical since the amount of logic required to sum up all the partial products is higher than the 8- and 16-bit cases.
Hence, sharing the multiplication resources among all the different bitwidth "regions", or even only sharing the adder tree to sum up over all the partial multiplication contributions, would be detrimental from the timing viewpoint: the additional combinatorial logic to select, split and distribute the operands and to enable the selected bit-width SIMD operation would have a negative impact on the overall speed.
The main drawback of our choice is in terms of area since we replicate hardware resources. 
As a direct consequence, the power consumption of the core system suffers a slight increase as well. To mitigate this effect on power consumption, we add a set of registers on the inputs of each bit-width region, and we perform clock gating to avoid switching for operands not involved in the current SIMD operation.

Despite the area overhead of 19.9\% with respect to the baseline Dot-Product unit, the extended unit does not increase the critical path of the system, and it does not require pipeline stages in between the multiplication and the accumulation phases.
Pipeline registers would result in execution stalls when computing back-to-back operations, introducing a huge overhead to the QNN workload, where most of the computation consists of sum-of-dot-product operations.
Moreover, the power consumption of the core is kept unaltered thanks to our power-aware design, as shown in Section \ref{sec:impl_results}.

\subsection{Compute\&Update Instruction}
\label{sec:compute_update}
%
In this section, we propose our hardware solution to further increase the speed-up of the QNN workload on RISC-V based pipelines.
To perform a MAC operation or a SIMD \textit{dot product} instruction on a RISC-based in-order single-issue processor, we first need to bring the two operands involved in the computation into the RF, at the cost of two load operations.
This means that only one-third of the executed instructions are relevant to the computation itself (i.e., the MAC instruction). We can formalize the concept defining the MAC operation efficiency (OPEF) metric that, in the case highlighted before, is equal to $0.33$.

Since most of the QNN workload consists of MAC operations, we want that the OPEF is as high as possible to achieve high performance and efficiency, knowing that it cannot be higher than one on a single-issue processor (by construction).
Data reuse at the RF level is an effective strategy to increase the OPEF of the MAC computation, as reported in \cite{garofalo2019pulp} and already discussed in Section \ref{sec:QNN_EXEC_MODEL}.
The innermost loop of the \textit{MatMul} kernel of PULP-NN (Fig.~\ref{fig:matmul_pulpnn}) reuses two activations in the RF over 4 filters.
This layout reduces the cost of the \textit{sdotp} operations down to only six loads, bringing the OPEF to $0.57$, with an improvement of 1.72 $\times$ with respect to the baseline.
%
%

Our solution to improve the MAC efficiency even more without giving up the flexibility of a general-purpose RISC-V processor consists of the architectural and micro-architectural design of Mac\&Load instructions, aiming at an OPEF close to 1.
%
We explore two different designs of the Mac\&Load operations for integration in \textit{XpulpNN} and discuss their respective benefits and the drawbacks, aiming at the best trade-off between performance and implementation costs in terms of area, timing, and power consumption. 
To introduce the intuition at the basis of the Mac\&Load paradigm, we discuss the assembly code of the \textit{MatMul} kernel reported in Figure~\ref{fig:matmul_pulpnn}.(b).
%
%
To hide the overhead of load operations, we propose to fuse the inner loop SIMD MAC (\textit{pv.sdotp}) with the load within a single Mac\&Load instruction. This is possible since the increment value (one word) is the same for all iterations, so it can be hardwired into the micro-architecture without being encoded into the instruction itself.
\begin{figure*}
  \centering
  \subfigure{\includegraphics[width=.45\linewidth]{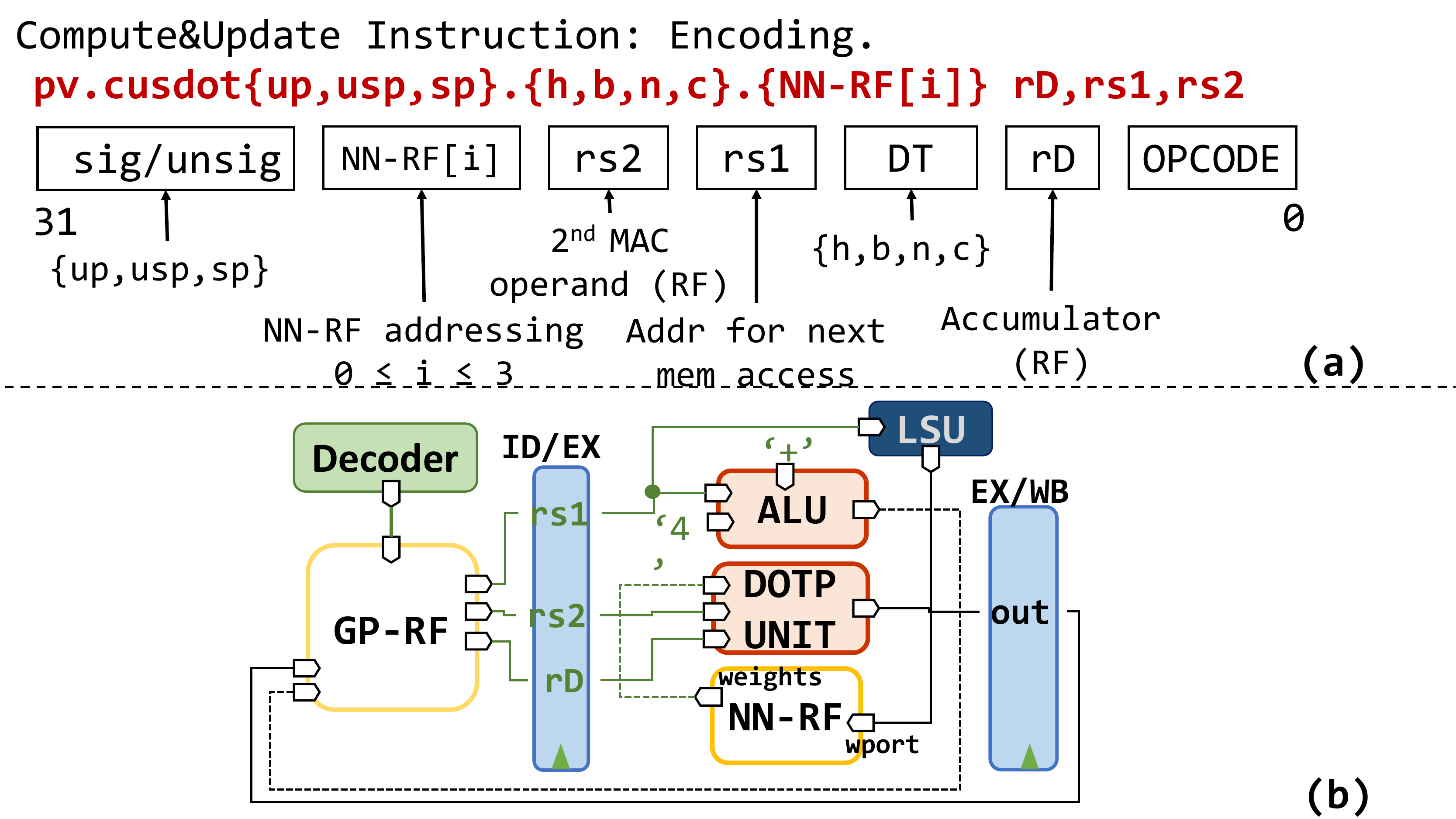} }%
    \qquad 
    \subfigure{\includegraphics[width=.45\linewidth]{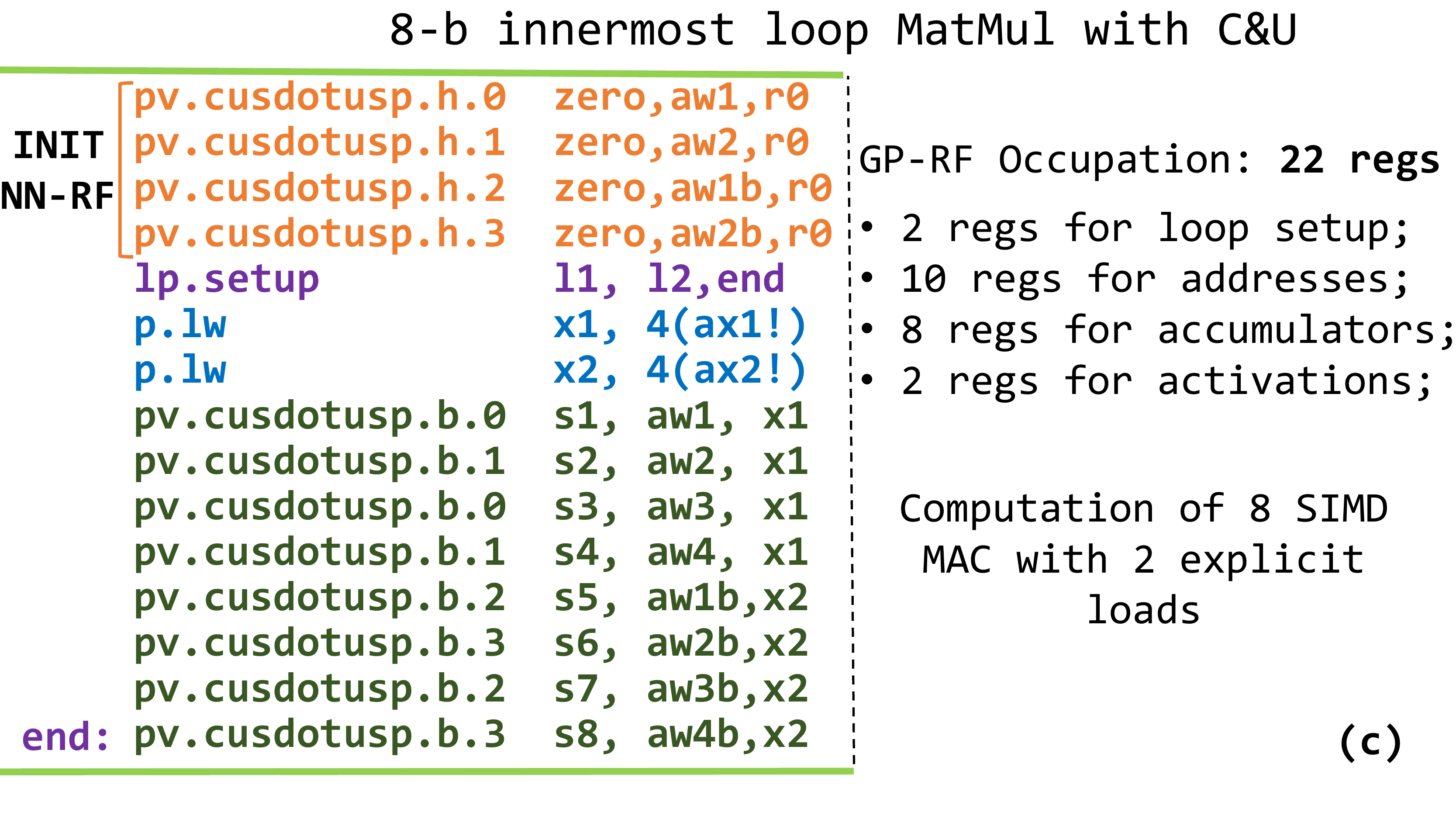} }%
  \caption{In (a) the prototype of the Compute\&Update (C\&U) instruction is reported: the MSBs encode the interpretation of the operands, ``NN-RF[i]'' selects the current NN-RF register, ``rs1'' is the address for the next memory access, ``rs2'' is the second operand for the MAC unit, while DT encodes the data type of the operands (symmetric) and ``rD'' is the accumulator. In (b) we see the datapath to enable the C\&U instruction. We add the NN-RF with one write port (connected to the LSU that fetches the new data accordingly to the ``rs1'' address) and one read port (multiplexed with the operand coming from the GP-RF) to feed the DOTP Unit. The ALU accepts the ``rs1'' operand to increment it by one word (`` $+4$'') and store it back to GP-RF. (c) depicts the innermost loop of the \textit{MatMul} kernel. Before the loop, we need extra instructions to initialize the dedicated NN-RF registers that do not affect the performance. Inside the loop we occupy in total 22 regs of the GP-RF and reduce the load costs for the MAC down to 2 operations, bringing the OPEF to 0.8. }
  \label{fig:compute_update}
\end{figure*}

%
In the first design of the Mac\&Load instruction, which we called Compute\&Update (``C\&U''), one of the operands of the Dotp Unit (e.g., one of the weights) is updated with a new memory element from the Load-Store Unit (LSU) of the core as soon as the SIMD MAC operation consumes it. The LSU accesses the memory location indicated by the ``rs1'' operand, as depicted in Figures \ref{fig:compute_update}.(a) and \ref{fig:compute_update}.(b). Afterward, the address consumed by the LSU is updated by one word in the ALU and stored back into the RF, similarly to the post-increment load of the \textit{XpulpV2} ISA \cite{gautschi2017near}. Data hazards, if any, are handled by stalling the pipeline exploiting the same signals of normal load instructions.
%
%

The RI5CY general-purpose RF (GP-RF) has two write ports, but the C\&U instruction requires three accesses to store the output of the \textit{dotp} operation, the updated address, and the new memory element. To avoid an additional cycle of latency, we would need to extend the GP-RF with one additional write port, which would be too expensive in terms of power and area.
Our lightweight solution is therefore to provide the EX-stage of the core with a very small register file dedicated to this computation paradigm, namely the Neural Network Register File (NN-RF, as visible in Figure~\ref{fig:compute_update}.(b)). The NN-RF is provided with one read port to feed the MAC unit with one operand and one write port to receive a new data word coming from the memory through the LSU. The NN-RF is sized in a way that all the loads related to the update of the weights in the innermost loop of the \textit{MatMul} kernel are masked. From our exploration, the optimal number of registers is 4.

As visible from Figure~\ref{fig:compute_update}.(a), the addressing of the NN-RF registers (``NN-RF[i]'' field) is hard-encoded into the instruction to compress as much as possible all the necessary information to execute the C\&U in the 32-bits of the encoding space. This causes the addition of four different C\&U instructions, each one controlling one register of the NN-RF.
%
We added support for a C\&U version of all the \textit{sdotp} based instructions, interpreting the operands as signed/unsigned-signed/unsigned (\textit{{sp,usp,up}}) and supporting 16-bit down to 2-bit SIMD operands (\textit{{h,b,n,c}}).

To enable the MAC computation with one operand coming from the NN-RF, the Dot-Product unit is further modified by multiplexing its first operand coming from the GP-RF with the read port of the NN-RF (see Figure~\ref{fig:compute_update}.(b)). Anytime the C\&U instruction is issued in the EX-stage, the Dotp-Unit fetches its first operand (the weight element in the case of the PULP-NN \textit{MatMul}) from the NN-RF.
This micro-architecture enables the execution of the C\&U instruction in one clock cycle of latency when the pipeline is fully operative and no stalls occur on the LSU-memory interface. 

By replacing the \textit{pv.sdotusp} instructions with the C\&U equivalents in the innermost loop of the \textit{MatMul} kernel, we are able to reduce the costs of explicit loads down to 2 with 8 SIMD MAC operations, as reported in Figure~\ref{fig:compute_update}.(c) where we take as an example an 8-bit kernel.
More in depth, we need some instructions of initialization to fill the NN-RF registers with the first operands involved in the MAC computations inside the loop. These few extra instructions do not affect the performance since they lay outside the critical loop.
This implementation of the \textit{MatMul} increases the OPEF to $0.8$, further gaining a $1.40\times$ of improvement with respect to the original PULP-NN solution.

Despite the efficiency improvement achieved, we noticed some limitations relate to the C\&U operation. The main drawback is that we need to update the NN-RF register consumed with the MAC operation at each instruction execution. This is not a concern from a functional point of view since we are always able to mask all non-necessary loads into the fused instruction. 
However, the load operations are performed by the Load unit of the core, causing energy-expensive accesses to the memory and interconnect.
In the context of tightly coupled shared-memory clusters, these additional loads create unnecessary contention, which degrades the overall performance.

Moreover, due to this ``context-based'' dependency, in the \textit{MatMul} we need to use two different registers of the GP-RF to address the same weight location in the memory. If we refer to Figure~\ref{fig:compute_update}.(c), the ``aw1'' address will be incremented by the \textit{pv.cusdotusp.b.0} instruction by one word to fetch the next weight from the memory. The consumed and discarded weight is also needed in the computation with the ``x2'' activation element. To fetch the correct weight again, we must occupy another register, namely ``aw1b''.

The weakness is that we are not exploiting data locality on the weight elements anymore, and we are occupying redundant registers into the GP-RF. The number of occupied registers remains unchanged with respect to the \textit{MatMul} of the PULP-NN library. Hence, also in this case, it is not possible to exploit the ``$4\times4$'' \textit{MatMul} data layout and its superior data reuse characteristics. 

\subsection{NN Sum-of-Dot-Product Instruction}
\begin{figure*}[t]
  \centering
  \subfigure{\includegraphics[width=.45\linewidth]{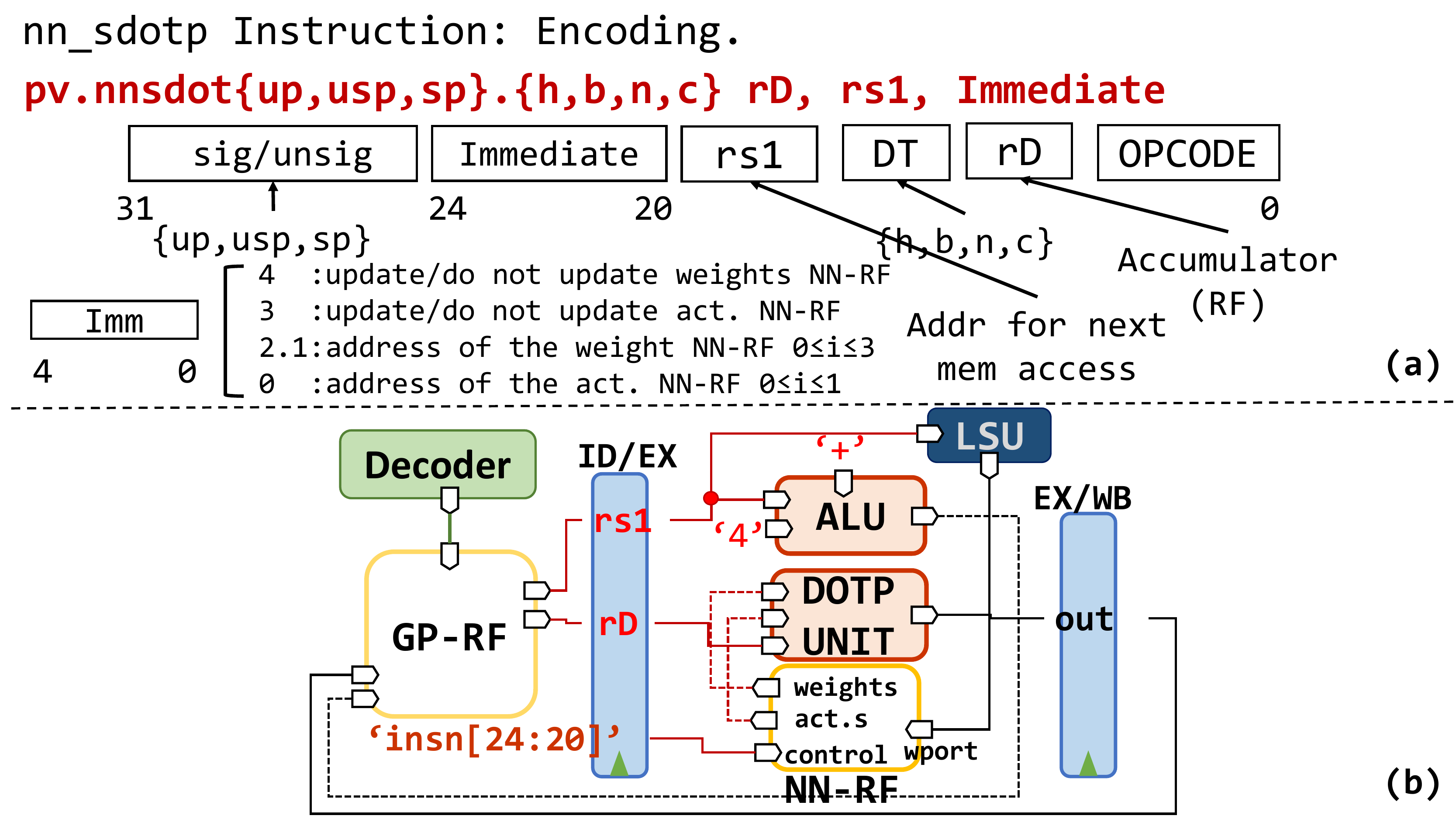} }%
    \qquad 
    \subfigure{\includegraphics[width=.45\linewidth]{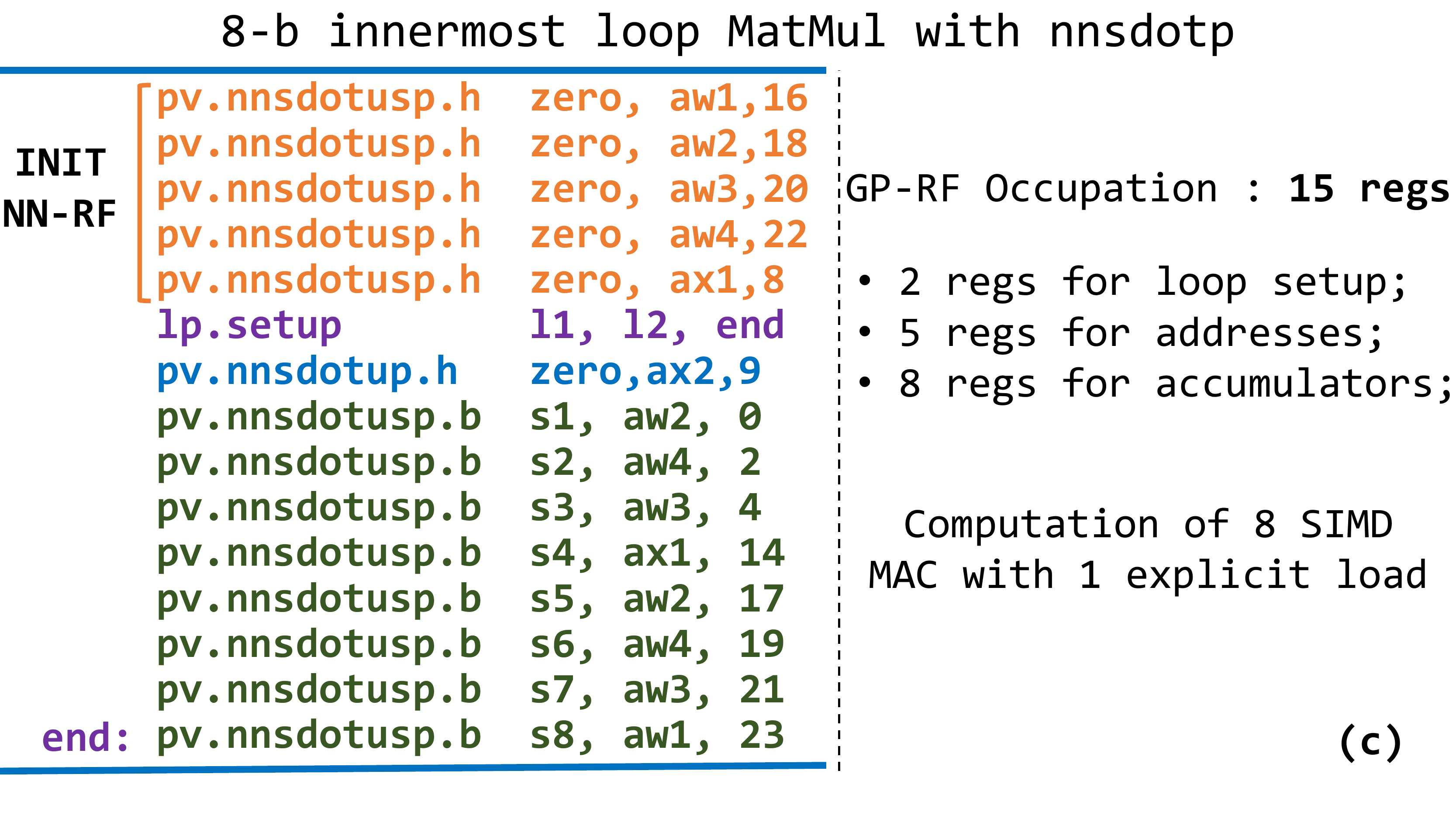} }%
  \caption{In (a) the Figure reports the encoding of the \textit{nn\_sdotp} instruction and describes the Immediate field. In (b) we depict the micro-architecture design to support the instruction in the RI5CY pipeline whereas (c) shows the \textit{MatMul} innermost loop implemented with the \textit{nn\_sdotp} instruction, highlighting the resources of the GP-RF occupied.}
  \label{fig:nn_sdotp}
\end{figure*}

The alternative version of the Mac\&Load instruction we propose, namely ``nn\_\textit{sum-of-dot-product} (nnsdotp)'', overcomes the flexibility issues of the C\&U presented above but requires more hardware resources to be integrated with the micro-architecture of the core.
More in detail, we provide a solution that allows the operands stored into the NN-RF to be kept there as long as needed before being updated with the load operation of the fused \textit{nnsdotp} instruction. 
This reduces the memory traffic, allows a higher grade of flexibility for data reuse (we are not limited by the compiler scheduler on the time we can keep an operand into the GP-RF), and solves the problem of using two different registers to encode the same address.
The drawback of the \textit{nnsdotp} is that the encoding of the new instruction is more complex. The functionality described above is encoded in a 5-bit Immediate field. This reduces the number of bits available to address another register of the GP-RF to feed the MAC unit with the second operand. Due to the regular structure of the \textit{MatMul} though, this is not a concern at all. Rather, we can extend the NN-RF with two additional registers to host the two activation elements involved in the innermost loop computation of the \textit{MatMul}. 
At the cost of a larger NN-RF compared to the solution adopted with the C\&U instruction, this solution guarantees more flexibility and performance.

As visible in Figure~\ref{fig:nn_sdotp}.(a), the 5-bit immediate addresses the NN-RF operands to be used in the current MAC operation: Bit 0 selects the activation register, bit 1\&2 select the weight register, and bits 3\&4 are set when we want to update either the addressed activation register or the weight register, respectively. Since we cannot update both weight and activation registers concurrently having a single LSU, these bits of the Immediate are mutually exclusive.
To support this mechanism in hardware (see Figure~\ref{fig:nn_sdotp}.(b)), we provide the NN-RF with an additional read port that is multiplexed with the operand coming from the GP-RF to feed the Dotp Unit, as described above. Only when the \textit{nnsdotp} instruction is issued, the Dotp Unit will receive both input operands from the NN-RF.
The immediate bits act as control signals for the NN-RF. 

The hardware cost of the \textit{nnsdotp} instruction consists of the additional NN-RF with one write, two read ports, and some logic to distribute the operands to the Dotp-Unit. The arithmetic blocks are already present in the micro-architecture. Hence, the impact of both the Mac\&Load instructions proposed is negligible in terms of the maximum frequency of the RI5CY core.
From a power consumption point of view, the \textit{nnsdotp} implementation has a non-negligible impact due to the additional NN-RF with two read ports and one write port. To avoid unnecessary switching activity when the \textit{nnsdotp} is not executed, we perform operand isolation on the critical operands (e.g., at the input of the multiplexers of the Dotp Unit), and perform clock gating in the NN-RF block.

The implementation of the \textit{MatMul} kernel using the \textit{nnsdotp} instructions is reported in Figure~\ref{fig:nn_sdotp}.(c). Before entering the innermost loop of the \textit{MatMul} we need to initialize all the  NN-RF registers. In this case, contrarily to the previous kernel with the C\&U instruction, we pay only one explicit load instruction to perform the same number of \textit{dotp} instructions, increasing the OPEF up to $0.88$, with an improvement of 1.1$\times$ with respect to the C\&U case.

A major benefit of the kernel highlighted in Figure~\ref{fig:nn_sdotp}.(c) is that the occupancy of the GP-RF registers is reduced by 15 registers. This results by moving all the operands in the dedicated NN-RF, keeping the GP-RF free to host addresses for intermediate values and accumulators.
\begin{figure}[t]
    \centering
    \includegraphics[width=\linewidth]{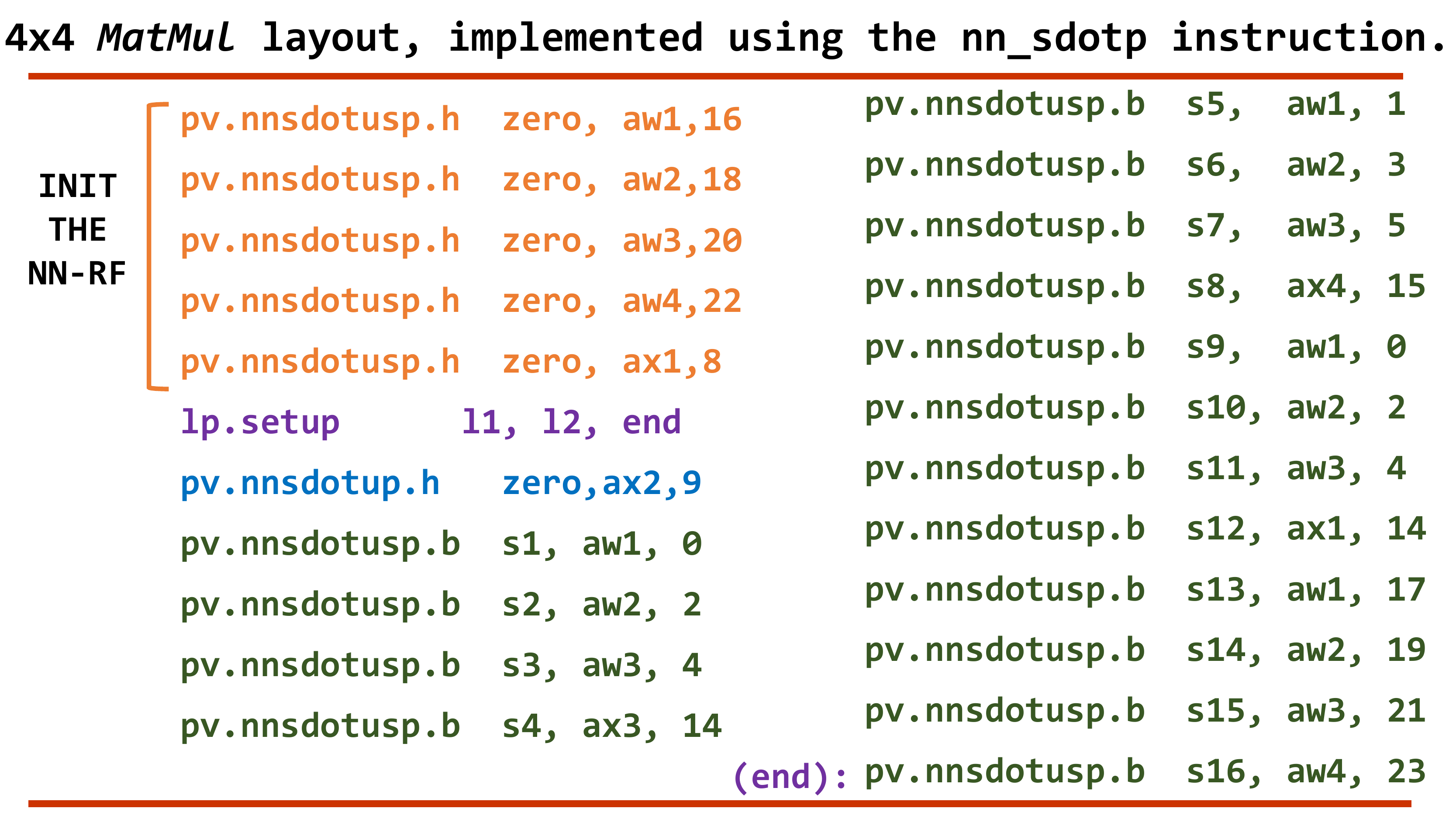}
    \caption{The figure shows the implementation of the ``4$\times4$'' \textit{MatMul} layout using the \textit{nn\_sdotp}. Storing the SIMD \textit{sdotp} operands into the NN-RF reduces the pressure on the GP-RF. More room is left to host more accumulators. The assembly code reported shows how the innermost loop of the \textit{MatMul} fit the register resources of the RI5CY core, thanks to the \textit{nn\_sdotp} instruction.}
    \label{fig:ml_4x4structure}
\end{figure}

This condition leaves space for the implementation of the ``$4\times4$`` \textit{MatMul} structure. We need to fetch two additional elements from \textit{im2col} memory buffers, whose addresses are stored into the GP-RF while the elements itself into the NN-RF. Reusing the weights also over the new activations, we can compute two additional pixels over four adjacent output channels (8 additional accumulators). Doing the math the occupancy of the GP-RF is of 32 registers (including the control registers for the HW loop), fitting the availability of the RI5CY GP-RF.
This intuition is demonstrated by the implementation of the ``$4\times4$'' kernel highlighted in Figure~\ref{fig:ml_4x4structure}. Following exactly the same strategy as in the other cases with the initialization of the NN-RF, we pay a single load instruction to execute 16 \textit{sdotp} operations, pushing the OPEF to 0.94, very close to the structural limit of 1.

\begin{figure}[t]
    \centering
    \includegraphics[width=\linewidth]{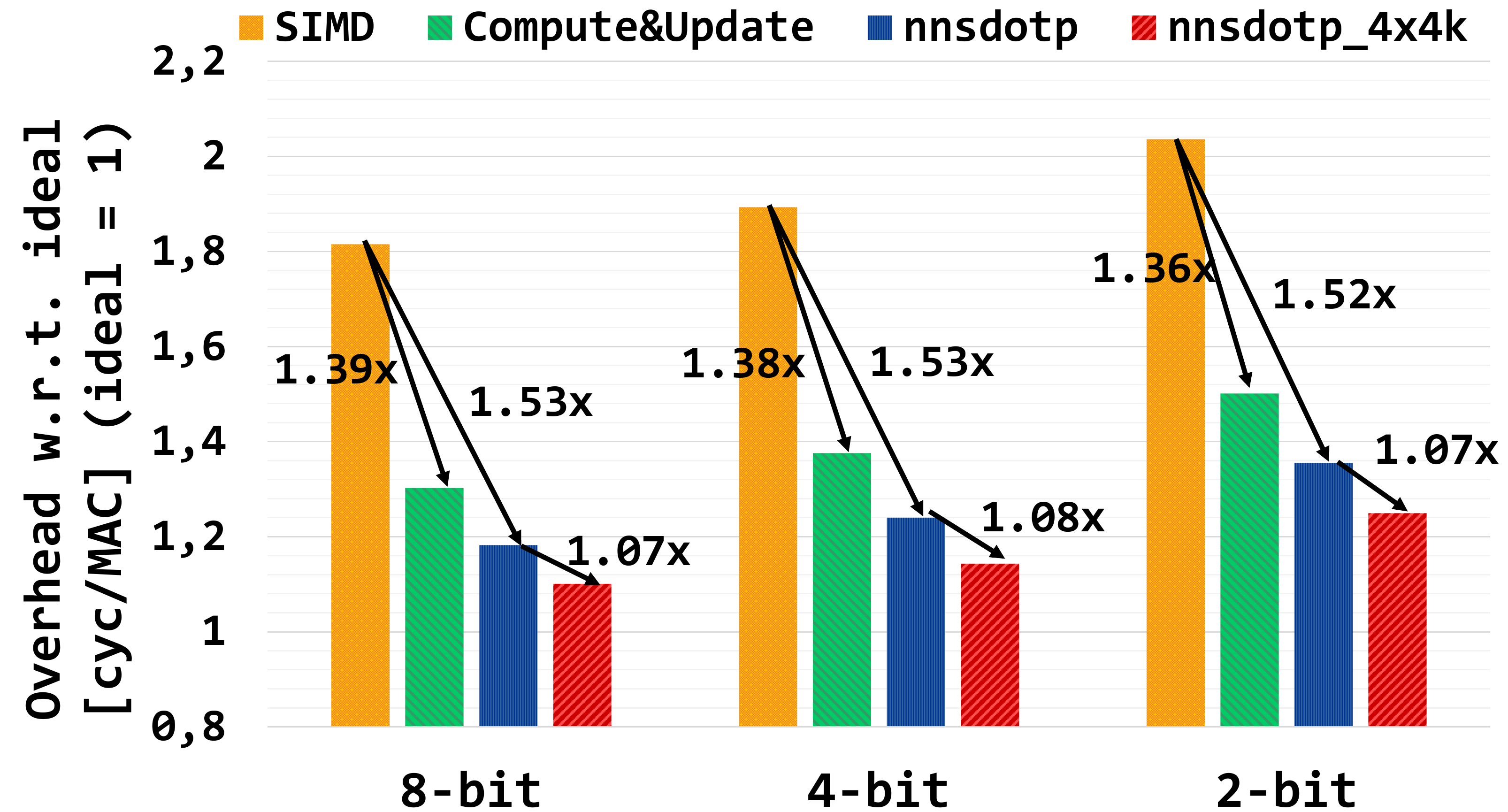}
    \caption{Invers of the Efficiency (lower is better) of the Matrix Multiplication kernel. The bar graph shows the cycles needed to the core to perform one SIMD MAC operation (4x8-bit. 8x4-bit, 16x2-bit respectively). The classical SIMD sdotp (XpulpNN) and the two versions of the MAC\&Load instructions (macload, nn\_custom) are considered. The nn\_custom allows also to enlarge the Matrix Multiplication layout (nn\_custom\_4x4k).}
    \label{fig:OPEF_1core}
\end{figure}

To assess the benefits of the mac\&load instructions at the micro-architecture level, we run simulations of the extended core executing multiple variants of the \textit{MatMul} kernel: first using only the SIMD operations (\textit{pv.sdotp}), and then using the C\&U instruction and the \textit{nn\_sdotp} operation. For the latter case, also the optimized kernel layout is considered. Figure~\ref{fig:OPEF_1core} reports the number of cycles required to perform a SIMD MAC operation (i.e., one \textit{dotp} 8-bit operation counts as one MAC).
As visible, the C\&U improves the efficiency by 1.39 $\times$ with respect to the SIMD case. Thanks to the enhanced \textit{nn\_sdotp} instruction, after initializing the NN-RF registers, the innermost loop of the \textit{MatMul} runs 1.10 $\times$ faster than in the C\&U case and 1.53$\times$ faster than the SIMD case. Finally, optimizing also the \textit{MatMul} layout, we gain an additional 1.07 $\times$ improvement with respect to the ``4$\times$2'' layout and the  \textit{nn\_sdotp}, with only 1.08 cyc/MAC, 1.65$\times$ higher than the SIMD case.

\section{XpulpNN Integration}
\subsection{Cluster integration}
\label{sec:clust_integration}
\begin{figure}[t]
    \centering
    \includegraphics[width=0.98\linewidth]{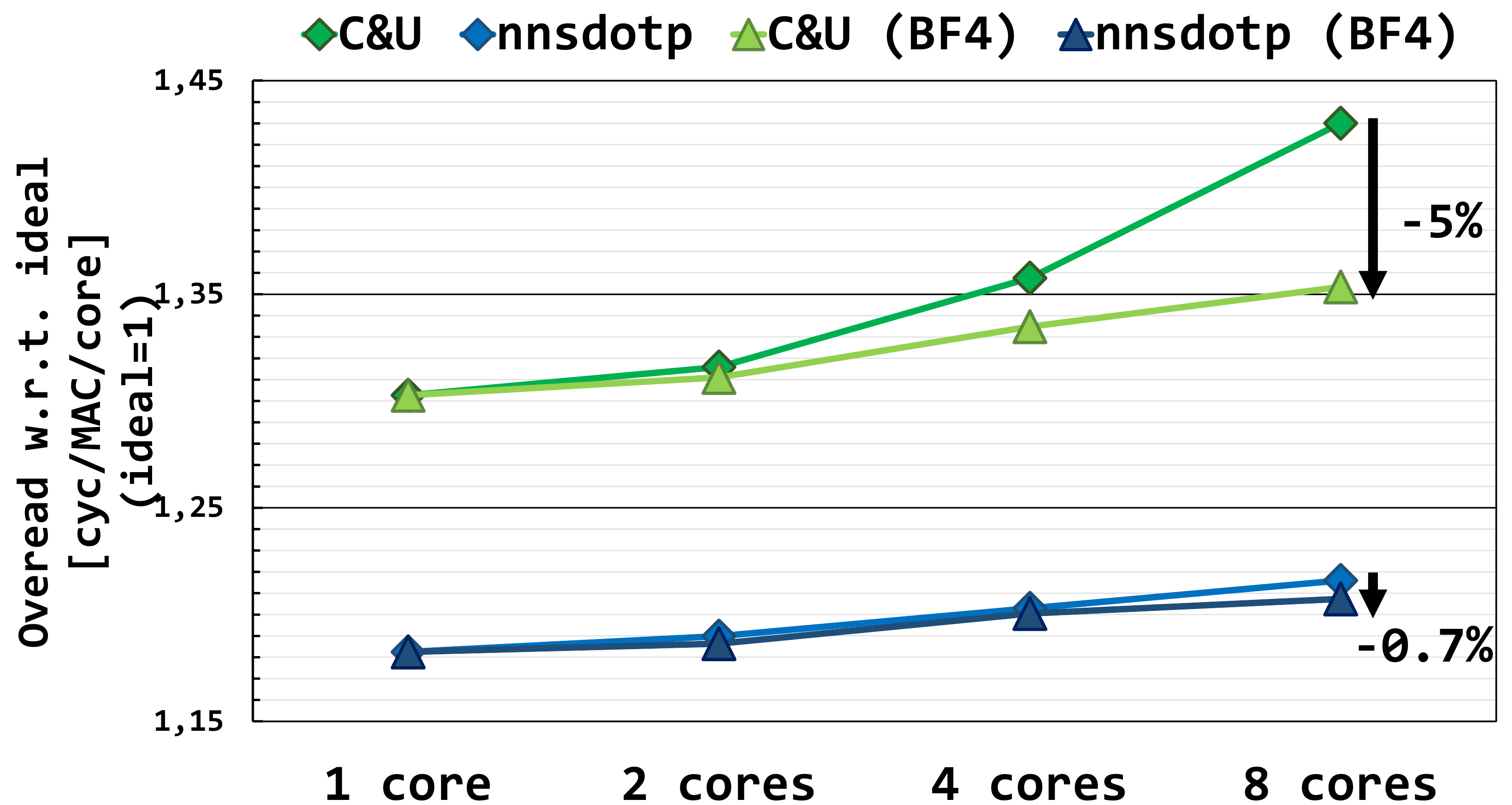}
    \caption{Inverse of the MAC Operation Efficiency (lower is better) of the PULP cluster on 8-bit Matrix Multiplication (\textit{MatMul}) kernels.}
    \label{fig:opef_multicore}
\end{figure}

After evaluating the improvement of the \textit{XpulpNN} ISA on a single-core execution of the \textit{MatMul} kernel, we integrate the extended RI5CY core into a PULP cluster of eight processors.
Since the QNN workload is highly parallelizable, we expect a near-linear scaling of the performance when moving from single- to multi-core contexts \cite{garofalo2020pulp}. 
We report in Figure~\ref{fig:opef_multicore} the results of the execution of the 8-bit \textit{MatMul} kernel in terms of cycle needed by each core to execute a SIMD MAC operation, considering the execution of the kernel first with the C\&U and then with the  \textit{nn\_sdotp} instruction.
The analysis carried out shows some drawbacks of the C\&U instruction that limits the efficiency of the computation in a multi-core context.
As visible from Figure~\ref{fig:opef_multicore}, when executing the \textit{MatMul} kernel with C\&U on eight cores, its efficiency decreases with respect to the single-core execution. As described in Section \ref{sec:compute_update}, the C\&U generates non-negligible traffic on the core-memory interface. This traffic results in many TCDM contentions in a multi-core context, causing each core to wait for the data from memory more than one cycle. 
Splitting the L1 memory over more banks, we are able to partially limit this effect. More in detail, if we consider a banking factor of four (``BF4'') (i.e., we double the baseline banking factor of two), the efficiency of the computation on eight cores increases by $5\%$, almost reaching the ones of the single core. However, this choice has a non-negligible impact on the power consumption of the system.
Instead, the \textit{nn\_sdotp} does not suffer from this limitation, thanks to its capability to keep in the NN-RF one operand as long as we need, reducing the traffic on the core-memory interface when not needed.
In a baseline configuration of the cluster (i.e., banking factor 2), the  \textit{nn\_sdotp} reaches almost the same efficiency as in the ``BF4'' configuration.


\subsection{Compiler \& Parallel Programming Support}
\label{sec:compiler}
All the instructions of the \textit{XpulpNN} ISA extensions can be inferred in the C code through the explicit invocation of built-in functions. In contrast with assembly inlining, this approach enables the lowering of built-ins into the high-level intermediate representation (IR) used by the compiler backend, allowing target-specific optimization passes to maximize the reuse of operands and efficiently schedule the instruction flow.
This mechanism is essential to model the accesses to NN-RF consequent to Compute\&Update semantic. Programmers do not have the visibility of the variables stored in NN-RF registers since their updates are hidden side effects from the C code perspective.
The backend IR associated with the built-ins maintains track of these relations, and optimization passes take them into account.

This approach, of course, restricts the flexibility for the average embedded system programmer.
However, our purpose is to expose the PULP-NN library functions as APIs. Practically, programmers never have to dig into a list of optimized low-level primitives, but they can select a library function (e.g., a convolution kernel). An example of this integration is in \cite{burrello2020dory}, where the backend library is integrated into a vertical QNNs deployment flow.

\section{Experimental Setup and Results}
\label{sec:EXPERIMENTAL_RESULTS}

In this section, we evaluate \textit{XpulpNN} both from a physical viewpoint, measuring and discussing the costs of the micro-architectural implementation in terms of area, power, and timing overheads with respect to the baseline RI5CY core and from a performance and energy efficiency perspective, comparing the execution of QNN workloads on top of the presented architectures with the State-of-the-Art Hardware and Software solutions.

%
To this purpose, we integrate both the RI5CY and the extended RI5CY cores into a Parallel Ultra-Low-Power (PULP) cluster of eight processors and perform a full implementation of the system in the Global Foundries 22nm FDX technology.
\begin{figure}[t]
    \centering
    \includegraphics[width=\linewidth]{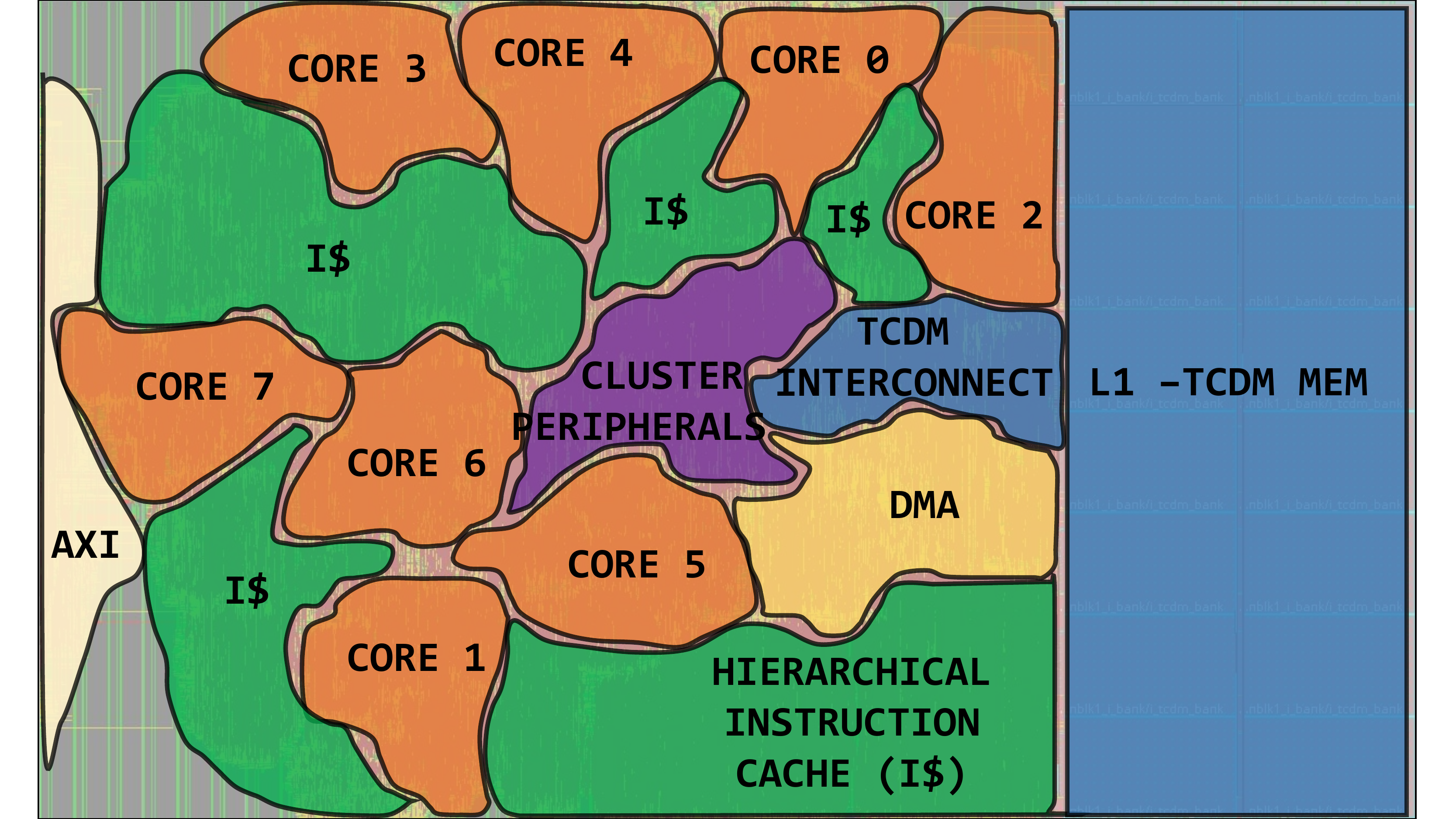}
    \caption{Placed and routed design of the PULP cluster with eight extended RISCY cores, supporting the \textit{XpulpNN} ISA.} 
    \label{fig:pulp_layout}
\end{figure}
We synthesize the two clusters with Synopsys Design Compiler-2018.3, and we perform a full place \& route flow using Cadence Innovus 17.11, in the worst-case corner (SS, 0.59V,  $-40^ \circ$/  $125^\circ$). The floorplan of the cluster is reported in Figure \ref{fig:pulp_layout}.
The total area of the cluster and of the core and the timing results are obtained from layout measurements. 
To perform power overhead evaluations, we run timing-annotated post-layout simulations in the typical corner and in different operating points, targeting common QNN workloads as well as general-purpose applications. 
Thus, all the results presented in the following include the overheads (i.e., timing, area, power) caused by the clock tree implementation, accurate parasitic models extraction, cell sizing for setup fixing and delay buffers for hold fixing (neglecting these would cause significant underestimations in the clock tree dynamic power).

To compare our solution with the State-of-the-Art in terms of performance and energy efficiency, we benchmark a set of convolution layers. 
In the context of this work, we focus on the implementation of the PULP cluster since we target a parallel execution of the QNN workload. We assume then that the cluster is connected to a simulated micro-controller system that has the only duty of activating the cluster and hosts an L2 level of memory containing the application code.
Since our goal is to improve the computing efficiency of the core kernels of a QNN inference task, we choose the layers such that their parameters fit the L1 memory of our systems to avoid additional overhead due to the memory transfers.
However, the selected convolution layers are representative of the common tiles used in such types of devices to deploy QNN inference \cite{burrello2020dory}.
The benchmarked layers operate on a 16 $\times$ 16 $\times$ 32 input tensor with a filter size of 64 $\times$ 3$\times$3$\times$32 and on a 32$\times$32$\times$32 input tensor with a filter size of 64$\times$3$\times$3$\times$32 respectively.
As described in Section \ref{sec:QNN_EXEC_MODEL}, after the \textit{MatMul} kernel, the intermediate results are compressed back into the desired precision through batch-normalization and activation functions.

\subsection{Implementation Results}
\label{sec:impl_results}
\begin{table}[t]
\caption{Area and Power Consumption Results. We consider typical and worst case corners for each operating point (HV= 0.8V, LV=0.65V). List of corners used for implementation: HV\_TYP: TT, 25C, 0.80V; HV\_SS: SS, 125C/-40C, 0.72V; LV\_TYP: TT, 25C, 0.65V; LV\_SS: SS, 125C/-40C, 0.59V. We also use fast corners for hold fixing. In all corners we use all permutations of parasitics (CMIN/CMAX/RCMIN/RCMAX). \textbf{Corners used for power analysis}: HV OP: TT, 25C, 0.80V, 660 MHz. LV OP: TT, 25C, 0.65V, 450 MHz. }
\label{tab:area_power}
\centering
\begin{tabular}{lcccc}

\hline \hline

\multicolumn{5}{c}{\textbf{Maximum Frequency {[}MHz{]}}} \\ 
\multicolumn{5}{c}{\textbf{of the cluster with Ext. RI5CY cores}} \\ \hline
\multicolumn{1}{c|}{} & \multicolumn{1}{|c|}{\textbf{HV}} & \textbf{LV} & \multicolumn{1}{|c|}{\textbf{HV\_SS}} & \multicolumn{1}{|c}{\textbf{LV\_SS}} \\ \hline
\multicolumn{1}{c|}{\textbf{PULP Cluster}} & \multicolumn{1}{|c|}{660 } & 450 & \multicolumn{1}{|c|}{400 } & 200 \\
\hline \hline

 & \multicolumn{2}{|c|}{\textbf{RI5CY}} & \multicolumn{2}{|c}{\textbf{Ext. RI5CY }} \\
 &\multicolumn{2}{|c|}{\textbf{(baseline)}} & \multicolumn{2}{|c}{\textbf{(with nn\_sdotp)}}  \\ \hline \hline

 \multicolumn{5}{c}{\textbf{Area {[}um2{]} (Overhead vs. baseline {[}\%{]})}} \\ \hline 
 
 \textbf{Tot. Cluster} & \multicolumn{2}{|c|}{970856} & \multicolumn{2}{|c}{1011254 \textbf{(4.1 \%)}} \\ \hline
 
\textbf{Tot. Cluster } & \multicolumn{2}{|c|}{995210} & \multicolumn{2}{|c}{1053446 \textbf{( 5.9 \%)}} \\
\textbf{(32 tcdm banks)} & \multicolumn{2}{|c|}{} & \multicolumn{2}{|c}{} \\ \hline

\textbf{Total Core} & \multicolumn{2}{|c|}{35131} & \multicolumn{2}{|c}{41296 \textbf{(17.5 \%)}} \\ \hline

\textbf{EX-Stage} & \multicolumn{2}{|c|}{13385} & \multicolumn{2}{|c}{17744 (32.6 \%)} \\ \hline \hline

\multicolumn{5}{c}{\textbf{Power Consumption of the CORE {[}mW{]}}}  \\ 
\multicolumn{5}{c}{ \textbf{on an 8-b MatMul (Overhead vs. baseline {[}\%{]})}} \\ \hline
\multicolumn{1}{c|}{} & \multicolumn{1}{|c|}{\textbf{HV}} & \textbf{LV} & \multicolumn{1}{|c|}{\textbf{HV}} & \multicolumn{1}{|c}{\textbf{LV}} \\ \hline
\multicolumn{1}{c|}{\textbf{Leak. Power}} & \multicolumn{1}{|c|}{2.13} & \multicolumn{1}{|c|}{0.96} & 2.22 & \multicolumn{1}{|c}{0.99}\\
\multicolumn{1}{c|}{\textbf{Dyn. Power}} & \multicolumn{1}{|c|}{2.94} & \multicolumn{1}{|c|}{1.30} & 3.01 & \multicolumn{1}{|c}{1.32}  \\
\multicolumn{1}{c|}{\textbf{Tot. Power}} & \multicolumn{1}{|c|}{3.05} & \multicolumn{1}{|c|}{1.35} & 3.12 \textbf{(2.1\%)} & \multicolumn{1}{|c}{1.39 \textbf{(2.5\%)}}  \\ \hline \hline

\multicolumn{5}{c}{\textbf{Power Consumption of the CORE {[}mW{]}}}  \\ 
\multicolumn{5}{c}{ \textbf{on a GP-application (Overhead vs. baseline {[}\%{]})}} \\ \hline 
\multicolumn{1}{c|}{} & \multicolumn{1}{|c|}{\textbf{HV}} & \textbf{LV} & \multicolumn{1}{|c|}{\textbf{HV}} & \multicolumn{1}{|c}{\textbf{LV}} \\ \hline
\multicolumn{1}{c|}{\textbf{Leak. Power}} & \multicolumn{1}{|c|}{0.108} & 0.055 & \multicolumn{1}{|c|}{ 0.122 } & 0.065 \\ \hline 
\multicolumn{1}{c|}{\textbf{Dyn. Power}}& \multicolumn{1}{|c|}{1.73}  & 0.76 & \multicolumn{1}{|c|}{1.76 (1.7\%)} &  0.78 (2.6\%) \\  \hline 
\multicolumn{1}{c|}{\textbf{Tot. Power}}& \multicolumn{1}{|c|}{1.84} & 0.82 & \multicolumn{1}{|c|}{1.88 \textbf{(2.17\%)}} & 0.85 \textbf{(3.7\%)}  \\ \hline \hline

\multicolumn{5}{c}{\textbf{Total Power Consumption of the PULP cluster {[}mW{]}}} \\ 
\multicolumn{5}{c}{\textbf{(Overhead vs baseline {[}\%{]})} }\\ \hline
\multicolumn{1}{c|}{} & \multicolumn{1}{|c|}{\textbf{HV}} & \textbf{LV} & \multicolumn{1}{|c|}{\textbf{HV}} & \multicolumn{1}{|c}{\textbf{LV}} \\ \hline

\multicolumn{1}{c|}{\textbf{MatMul 8-bit}} & 41.8 & \multicolumn{1}{|c|}{19.3} & 41.6 & \multicolumn{1}{|c}{19.3 (0.02 \%)} \\ \hline
\multicolumn{1}{c|}{\textbf{(with nn\_sdotp)} }& -- & \multicolumn{1}{|c|}{--} & 43.7 (5.11 \%) & \multicolumn{1}{|c}{21.5 (11.5\%)}\\ \hline
\multicolumn{1}{c|}{\textbf{MatMul 4-bit}} & -- & \multicolumn{1}{|c|}{--} & 35 & \multicolumn{1}{|c}{16.1} \\ \hline
\multicolumn{1}{c|}{\textbf{(with nn\_sdotp)}} & -- & \multicolumn{1}{|c|}{--} & 41.2 & \multicolumn{1}{|c}{19} \\ \hline
\multicolumn{1}{c|}{\textbf{MatMul 2-bit}} & -- & \multicolumn{1}{|c|}{--} & 42.9 & \multicolumn{1}{|c}{19.1} \\ \hline
\multicolumn{1}{c|}{\textbf{(with nn\_sdotp)}} & -- & \multicolumn{1}{|c|}{--} & 48.9 & \multicolumn{1}{|c}{24.1}\\ \hline
\multicolumn{1}{c|}{\textbf{GP Application}} & 27.6 & \multicolumn{1}{|c|}{ 12.9 }& 28.3 (\textbf{2.4 \%} ) &\multicolumn{1}{|c}{ 13.3 (\textbf{3.1 \%})}\\ \hline
\end{tabular}
\end{table}
 \begin{figure}[t]
     \centering
     \includegraphics[width= 0.95 \linewidth]{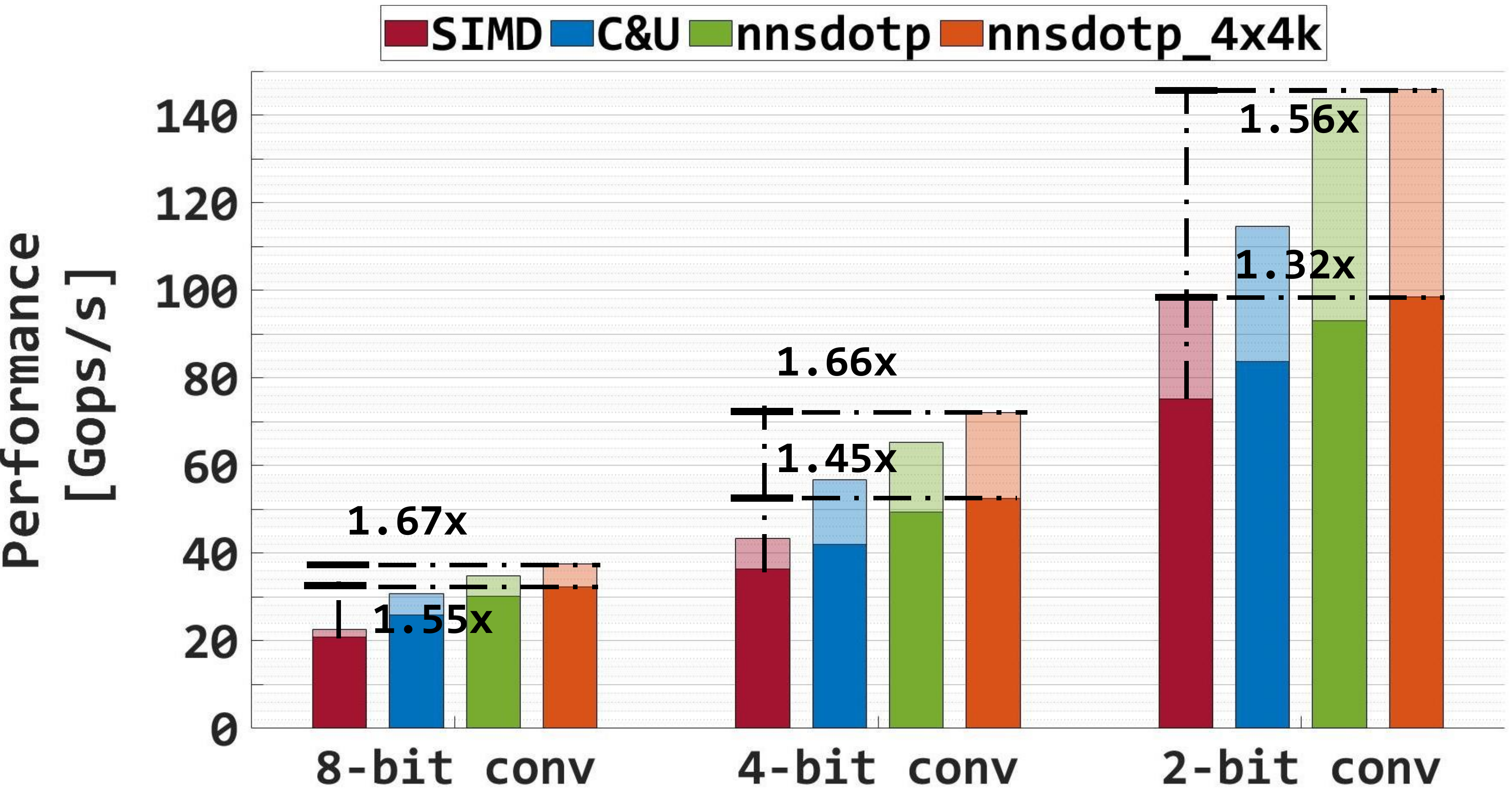}
     \caption{Performance of the 8 core PULP clusters over different bit-width precision Convolution kernels, implemented with the different instructions presented in this work. The lighted bars (higher-performance) refers to the \textit{MatMul} kernel only, while the darker ones include also the quantization procedure (hence, the whole convolution). The cluster runs in the best performance operating point, at 660 MHz, 0.8V in the typical corner.  }
     \label{fig:perf_400}
 \end{figure}
Table~\ref{tab:area_power} shows a comparison between the RI5CY core and the extended RI5CY, implementing the \textit{XpulpNN} ISA (with the mac\&loadv2), in terms of area and power consumption, estimated on post-layout simulations of different applications.
The total area of the extended RI5CY is $0.041 mm^2$, with an overhead of $17.5\%$ with respect to our baseline. Such increment is mostly due to the addition of the multipliers in the Dotp-Unit of the baseline core and of the extra-registers to build the NN-RF.
The cluster area instead is of $1\;mm^2$ with the new core, $~4\%$ higher than the baseline. In Table~\ref{tab:area_power}, we take into account also the cluster implementation with a banking factor of four to highlight the cost of this exploration in terms of area overhead. The cost of doubling the banking factor results in an additional area overhead of $4.2\%$.
As introduced in Section \ref{sec:ISA}, the duplication of the hardware resources into the Dotp-Unit allows us not to affect the critical path of the system. The maximum frequency achievable by both considered cores (RI5CY and the extended RI5CY) is the same.

Despite a non-negligible area overhead, the power consumption of the core is not affected significantly, as well as the power of the whole cluster system. To provide an accurate power estimation of the cores and characterize the whole system-level power consumption, we conduce post-layout power simulations in two different voltage corners: the high-voltage corner (TT, 660 MHz, 0.80V) and the low-voltage one (TT, 450 MHz, 0.65V). We test 8-bit Dot-product based operations, the new nibble and crumb instructions, as well as the mac\&load in its final version (nn\_sdotp).
Each kernel considered in the comparison is compiled with an extended GCC 7.1 toolchain that supports both \textit{XpulpV2} and \textit{XpulpNN} extensions. The Value Change Dump (VCD) traces are generated with Mentor Modelsim 10.7b and analyzed by Synopsys Prime Time 2019.12 to extract the power numbers.
As visible in Table~\ref{tab:area_power}, thanks to the clock gating techniques and to the operands isolation and despite the bigger core area, the extended RI5CY core runs an 8-bit Matrix Multiplication kernel (both the cores are using the 8-bit SIMD arithmetic instructions of the \textit{XpulpV2} ISA) in almost the same power envelope of the baseline core, with a power overhead of only ~3\% in both considered corners.
The same reasoning applies if we consider a General Purpose application, consisting of a mixture of the plain RISC-V ISA (RV32IMC) instructions such as load/stores, arithmetic, and control operations. 
This achievement is also visible at the system level, comparing the PULP cluster power consumption, demonstrating the light-weighted nature of the ISA extensions proposed in this work, and furthermore showing that we do not jeopardize the energy efficiency of the core on general-purpose benchmarks.


\subsection{Benchmarking}
\label{sec:Benchmarking}

 \begin{figure}[t]
     \centering
     \includegraphics[width= 0.95 \linewidth]{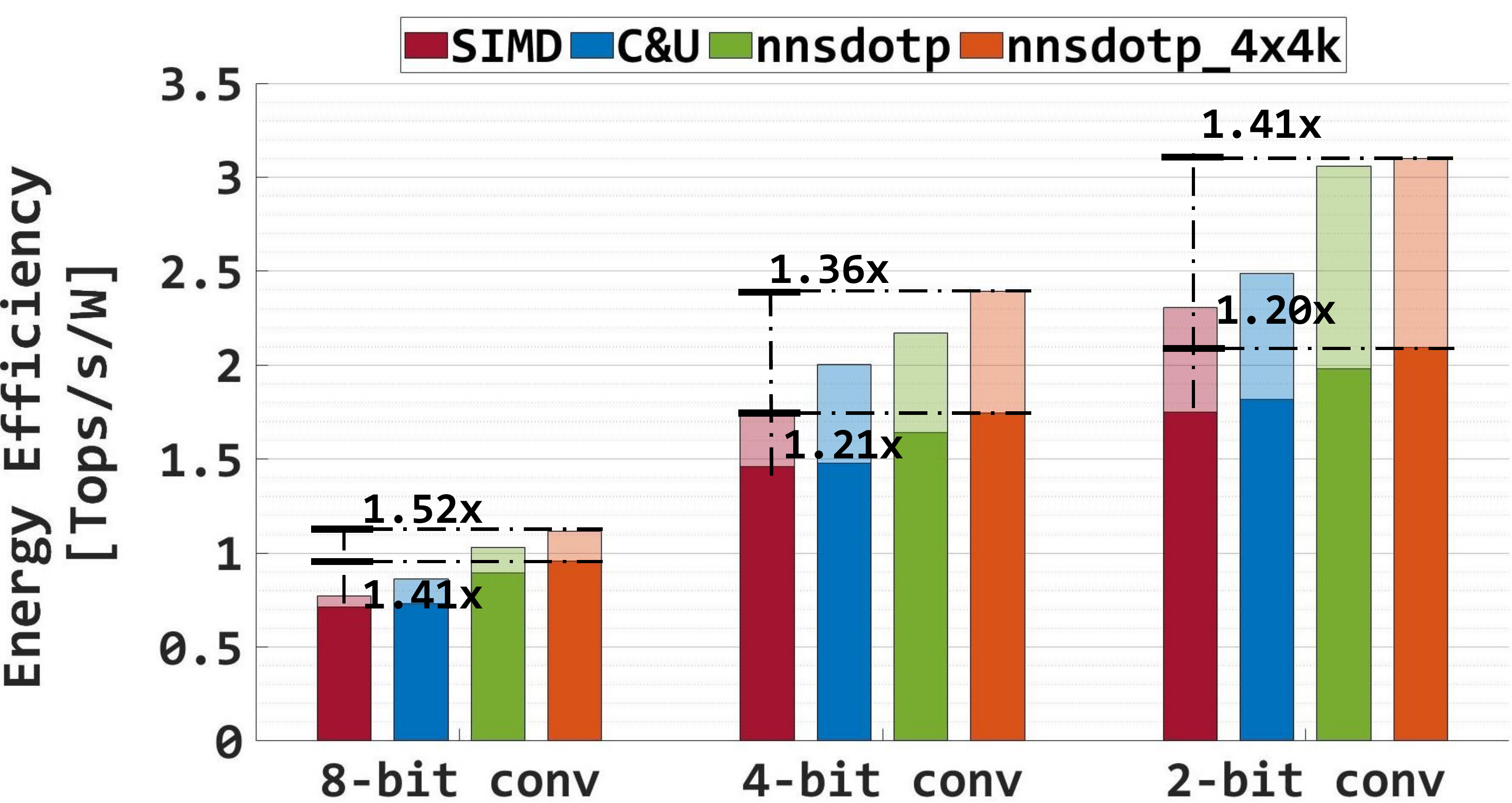}
     \caption{Energy efficiency of the convolutions on the 8 core PULP clusters. The graph compares the solutions described in this work.The lighted bars (higher-performance) refers to the \textit{MatMul} kernel only, while the darker ones include also the quantization procedure (hence, the whole convolution). The cluster runs in the best efficiency operating point, at 450 MHz, 0.65V, in the typical corner. }
     \label{fig:ee_400}
 \end{figure}

To evaluate the performance and the energy efficiency gain achieved with the proposed \textit{XpulpNN} extensions, we benchmark the convolution layers discussed above in different bit-width symmetric configurations (8-, 4-, and 2-bits). 
The kernels run on the extended RI5CY core, using different instructions of the \textit{XpulpNN} ISA: classical SIMD operations, compute\&update, nn\_sdotp and the nn\_sdotp optimizing the layout of the \textit{MatMul}.
This analysis aims at measuring the impact of the extensions on the whole convolution kernel of the PULP-NN library.
The performance achieved, as well as the energy efficiency, are measured at the high-voltage corner (TT, 0.8V, 25C) and the low-voltage corner (TT, 0.65V, 25C) respectively of the post-layout simulations  and reported in Figure~\ref{fig:perf_400} and \ref{fig:ee_400} respectively.
The peak performance and efficiency of the convolution layers are reached by implementing the \textit{MatMul} kernel with the nn\_sdotp instruction and an optimized 4$\times$4 layout. In the 8-bit case, the improvement with respect to the classical SIMD implementation of the \textit{MatMul} is 1.55$\times$ and $1.41\times$ in terms of performance and efficiency, respectively. The little degradation of these two metrics compared to the ideal case where we consider only the execution of the \textit{MatMul} kernel (bars in transparency in the Figure) is due to the quantization and compression of the intermediate \textit{MatMul} results.

The impact of the quantization is much higher on the 4- and 2-bit convolution layers, especially when we refer to the optimized \textit{MatMul} kernels. The reason for this behavior is that the computational cost for quantization does not depend on the bit-width of the compressed output feature map, meaning that it consists of the same operations no matter what is the precision of the final results. Considering the same layer parameters, the lower the precision of the \textit{MatMul}, the less the iterations of the innermost loop (since in one \textit{dotp} based operation we are actually performing 4, 8 or 16 effective MACs).
Hence, the effective improvements in the \textit{MatMul} kernel using the nn\_sdotp instruction are mitigated by the batch-normalization and activation step on 4- and 2-bit convolution layers. As visible from the Figure~\ref{fig:perf_400}, the performance improvement with respect to the classical SIMD implementation of the \textit{MatMul} passes from 1.66$\times$ (1.56$\times$) on the 4-bit (2-bit) \textit{MatMul} itself to 1.45$\times$ (1.32$\times$) on the whole 4-bit (2-bit) convolution layer.  
Obviously, these results directly translate into a corresponding degradation of energy efficiency. However, thanks to the optimized 4$\times$4 \textit{MatMul} kernel and the nn\_sdotp instruction, we boost the convolution efficiency by up to 1.41$\times$ with respect to the SIMD implementation.
\begin{figure}[t]
     \centering
     \includegraphics[width=0.9 \linewidth]{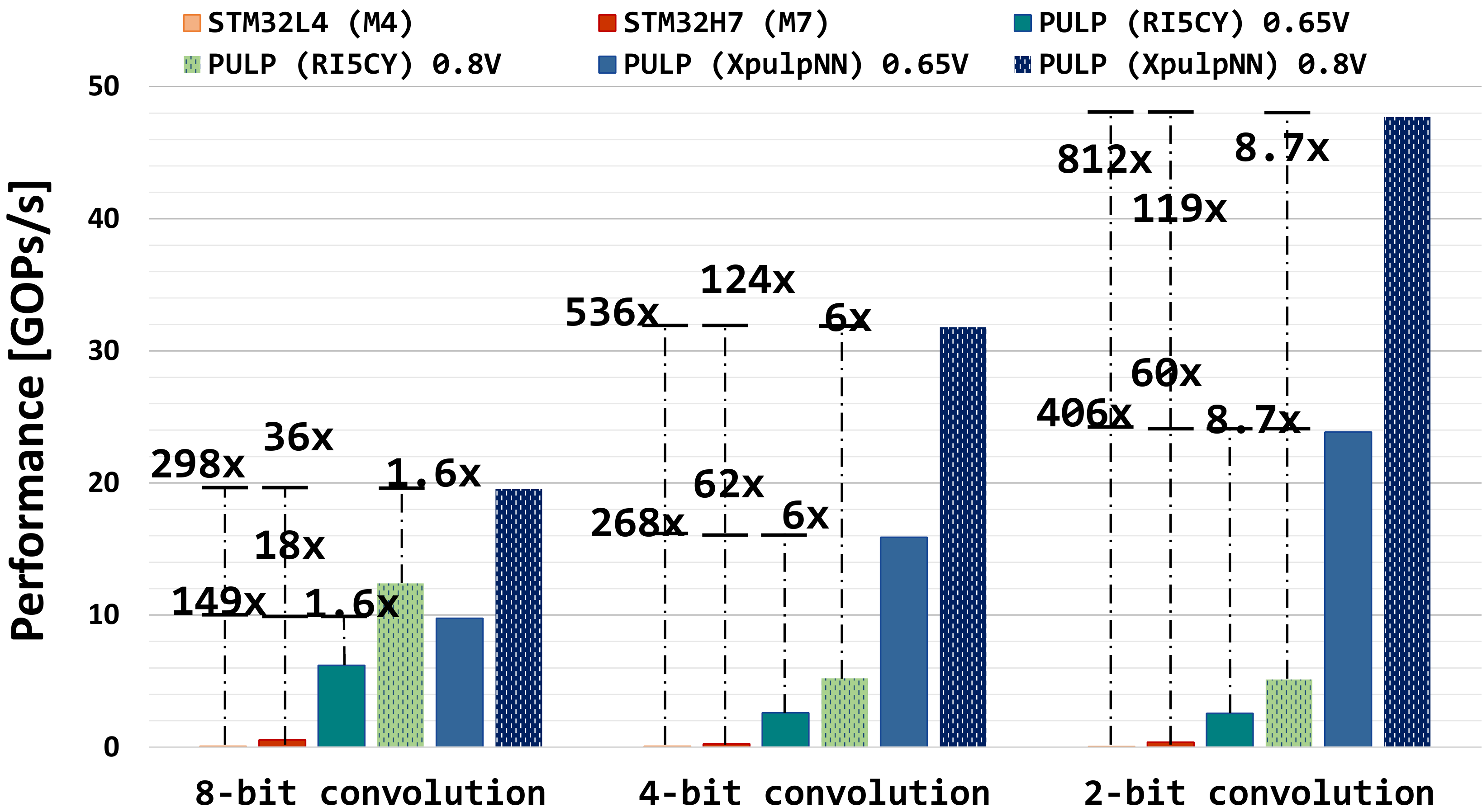}
     \caption{The Figure shows the comparison of this work with the State-of-the-Art (high-end STM32H7 and low-end STM32L4 MCUs) and with the baseline RI5CY cluter, in terms of performance. The PULP clusters run in two operating points: high-voltage (0.8V, 400 MHz) and low-voltage (0.65V, 200 MHz). 8-, 4- and 2-bit simmetric convolution kernels are benchmarked to carry out the comparison.}
     \label{fig:perf_SOA}
 \end{figure}
\begin{figure}[t]
     \centering
     \includegraphics[width=0.9 \linewidth]{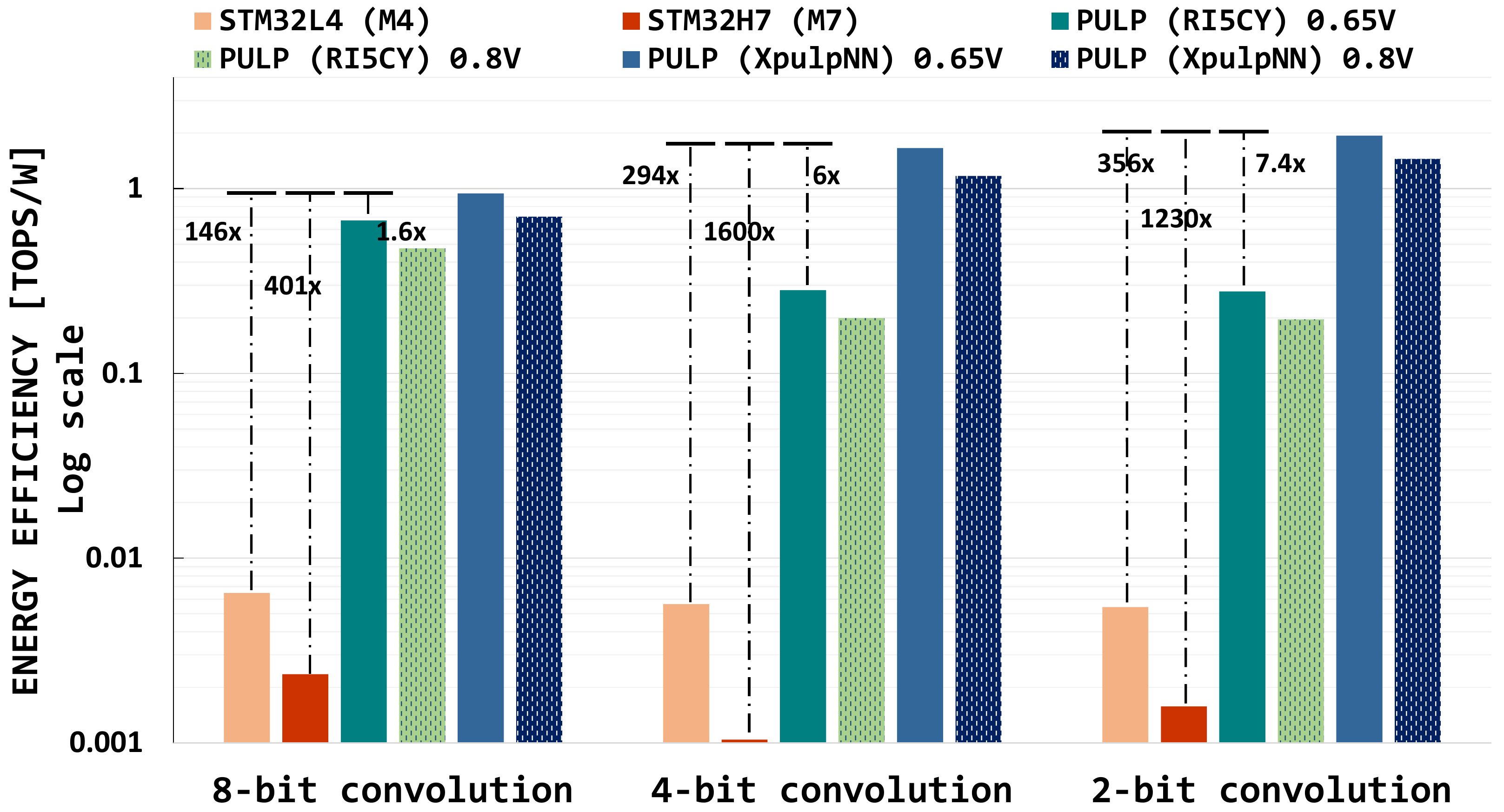}
     \caption{Energy efficiency comparison of the this work with State-of-the-Art and the baseline RI5CY clusters. 8-, 4- and 2-bit simmetric convolution kernels are benchmarked to carry out the comparison.}
     \label{fig:ee_SOA}
 \end{figure}

Despite the small degradation of performance and efficiency due to the quantization phases of sub-byte output activations, these cumulative improvements on the QNN kernels demonstrate the effective strategy of extending the ISA with domain-specific lightweight instructions to obtain high performance and energy efficiency on highly quantized  QNN kernels, without affecting the system on other domain applications efficiency. 
%

\subsection{Comparison With the State-of-the-Art}
\label{sec:Comparison}
To put our achievement in perspective, we compare our results with state-of-the-art existing hardware and software solutions in terms of performance and energy efficiency.
To carry out the comparison, we run the convolution layers on the RI5CY cluster using the PULP-NN library \cite{garofalo2020pulp} and on two off-the-shelf STM32H7  and STM32L4  commercial microcontrollers previously introduced in Section \ref{sec:related_work}, using the extended CMSIS-NN library \cite{rusci2018work}.
The performance and energy efficiency results are summarized in the Figure~\ref{fig:perf_SOA} and \ref{fig:ee_SOA} respectively.
For the implemented PULP cluster (with RI5CY and the extended RI5CY cores), we report two operating points: one at high-voltage, 0.8 V, 400 MHz and one at low-voltage, 0.65 V, 200 MHz, with the purpose to give insights on how much performance we trade-off with the energy efficiency at the highest voltage and vice versa.
It is important to note that, since the STM32 MCUs are commercial products signed-off in the SS corners, the power analysis of our solution is carried out in the SS operating points (i.e. considering 400 MHz (200MHz) as the frequency for the best performance (efficiency) points) for a fair comparison.
As visible from Figure~\ref{fig:perf_400}, with the same operating condition, we improve the performance of the 4-bit (2-bit) convolution layers by 6$\times$ (8.7$\times$) with respect to the RI5CY cluster. Thanks to the nn\_sdotp we are also able to increase by 1.6$\times$ the performance on 8-bit convolutions. Almost the same grade of improvement is reached on the energy efficiency of such kernels, demonstrating that both clusters run almost in the same power envelope despite the enhanced ISA and the additional hardware.
Also, the convolution kernels on the \textit{XpulpNN} PULP cluster at the high(low)-voltage operating point run from 298$\times$ to 812$\times$ (149$\times$ to 406$\times$) faster than the same kernels executing on the low-end STM32L4 using the CMSIS-NN library. In terms of energy efficiency, we outperform this microcontroller system by up to 356$\times$ in the best case (2-bit convolution, low-voltage operating point).
Our performance gain with respect to the high-end Cortex-M7 based STM32H7 microcontroller is more limited than the previous case, since the STM32H7 runs at 480 MHz and features a dual issue core. In this case, we outperform its performance by up to 119$\times$. Being a high-end microcontroller system the STM32H7 suffers in terms of energy efficiency, where we do better by up to three orders of magnitude, as visible in Figure~\ref{fig:ee_SOA}.

The presented results, coming out from the state-of-the-art comparison, are the consequence of the following insights: contrarily to ARM Cortex-M cores, the proposed solution has hardware support for 8-, 4- and 2-bit SIMD dotp-based operations and for the mac\&load instruction. The STM32 based systems consist of a single-core chip, while our target architecture is a computing cluster of eight processors to improve the efficiency of the computation. The remaining performance/efficiency is gained due to the more scaled technology used to implement the PULP cluster compared to the one of the STM32L4 (90nm) and of the STM32H7 (40nm).
In the end, the carried-out analysis shows for the first time that we can achieve ASIC-like energy efficiency on QNN workloads on fully programmable tiny MCU systems of the extreme-edge of the IoT. This outcome is obtainable by coupling the power-aware micro-architecture design and its integration in a multi-core computing cluster architecture with leading-edge near-threshold FDX technology.

%

\section{Conclusion}
Strongly quantized fixed-point arithmetic is considered the key direction to enable the inference of CNNs on low-power, resource-constrained edge devices.
However, the lack of hardware support for low-bitwidth arithmetic in the modern ISAs of MCU systems makes QNNs adoptions effective as a memory compression technique but not a solution to reduce time and energy required for computation. Besides, non-negligible overhead occurs to manipulate low-bitwidth data.

To overcome the limitations mentioned above, we have presented \textit{XpulpNN}, a set of low-bitwidth SIMD arithmetic extensions to the RISC-V ISA, which enables efficient computation of heavily QNN kernels at the extreme edge of IoT.
%
%
We have shown the benefits at the ISA levels of implementing QNN kernels with the new proposed instructions. Furthermore, we integrated the new extended RI5CY core in a multi-core computing cluster, showing a near-linear speedup of the performance compared to the single-core execution.

The implementation of the PULP cluster in leading-edge gf22 nm FD-SOI technology showed that: thanks to the power-aware micro-architecture design to support the new instructions, the \textit{XpulpNN} extensions do not jeopardize the efficiency of RI5CY on general-purpose applications; given the same technology, the energy efficiency on byte and sub-byte kernels has been improved by up to one order of magnitude with respect to the baseline RI5CY.

Our work shows at least two orders of magnitude improvements in performance and energy efficiency than state-of-the-art hardware and software solutions based on ARM Cortex-M cores.
This scenario paves the way to software programmable QNN inference at the extreme edge of the IoT, promising ASIC-like efficiency with higher flexibility.

\section*{Acknowledgment}
This work was supported in part by the EU Horizon 2020 Research and Innovation
projects OPRECOMP (Open trans-PREcision COMPuting, g.a. no. 732631) and
WiPLASH (Wireless Plasticity for Heterogeneous Massive Computer Architectures, g.a.
no. 863337) and by the ECSEL Horizon 2020 project AI4DI (Artificial Intelligence for
Digital Industry, g.a. no. 826060).

\ifCLASSOPTIONcaptionsoff
  \newpage
\fi



%
\bibliographystyle{IEEEtran}
\bibliography{IEEEfull,bibliography_xpulpnn.bib}

\begin{thebibliography}{10}
\providecommand{\url}[1]{#1}
\csname url@samestyle\endcsname
\providecommand{\newblock}{\relax}
\providecommand{\bibinfo}[2]{#2}
\providecommand{\BIBentrySTDinterwordspacing}{\spaceskip=0pt\relax}
\providecommand{\BIBentryALTinterwordstretchfactor}{4}
\providecommand{\BIBentryALTinterwordspacing}{\spaceskip=\fontdimen2\font plus
\BIBentryALTinterwordstretchfactor\fontdimen3\font minus
  \fontdimen4\font\relax}
\providecommand{\BIBforeignlanguage}[2]{{%
\expandafter\ifx\csname l@#1\endcsname\relax
\typeout{** WARNING: IEEEtran.bst: No hyphenation pattern has been}%
\typeout{** loaded for the language `#1'. Using the pattern for}%
\typeout{** the default language instead.}%
\else
\language=\csname l@#1\endcsname
\fi
#2}}
\providecommand{\BIBdecl}{\relax}
\BIBdecl

\bibitem{hassanalieragh2015health}
M.~Hassanalieragh, A.~Page, T.~Soyata, G.~Sharma, M.~Aktas, G.~Mateos,
  B.~Kantarci, and S.~Andreescu, ``Health monitoring and management using
  internet-of-things (iot) sensing with cloud-based processing: Opportunities
  and challenges,'' in \emph{2015 IEEE International Conference on Services
  Computing}.\hskip 1em plus 0.5em minus 0.4em\relax IEEE, 2015, pp. 285--292.

\bibitem{tokognon2017structural}
C.~A. Tokognon, B.~Gao, G.~Y. Tian, and Y.~Yan, ``Structural health monitoring
  framework based on internet of things: A survey,'' \emph{IEEE Internet of
  Things Journal}, vol.~4, no.~3, pp. 619--635, 2017.

\bibitem{li20185g}
S.~Li, L.~Da~Xu, and S.~Zhao, ``5g internet of things: A survey,''
  \emph{Journal of Industrial Information Integration}, vol.~10, pp. 1--9,
  2018.

\bibitem{palossi201964}
D.~Palossi, A.~Loquercio, F.~Conti, E.~Flamand, D.~Scaramuzza, and L.~Benini,
  ``A 64-mw dnn-based visual navigation engine for autonomous nano-drones,''
  \emph{IEEE Internet of Things Journal}, vol.~6, no.~5, pp. 8357--8371, 2019.

\bibitem{shi2016edge}
W.~Shi, J.~Cao, Q.~Zhang, Y.~Li, and L.~Xu, ``Edge computing: Vision and
  challenges,'' \emph{IEEE internet of things journal}, vol.~3, no.~5, pp.
  637--646, 2016.

\bibitem{reuther2019survey}
A.~Reuther, P.~Michaleas, M.~Jones, V.~Gadepally, S.~Samsi, and J.~Kepner,
  ``Survey and benchmarking of machine learning accelerators,'' \emph{arXiv
  preprint arXiv:1908.11348}, 2019.

\bibitem{daly2020through}
D.~C. Daly, L.~C. Fujino, and K.~C. Smith, ``Through the looking glass-2020
  edition: Trends in solid-state circuits from isscc,'' \emph{IEEE Solid-State
  Circuits Magazine}, vol.~12, no.~1, pp. 8--24, 2020.

\bibitem{hubara2017quantized}
I.~Hubara, M.~Courbariaux, D.~Soudry, R.~El-Yaniv, and Y.~Bengio, ``Quantized
  neural networks: Training neural networks with low precision weights and
  activations,'' \emph{The Journal of Machine Learning Research}, vol.~18,
  no.~1, pp. 6869--6898, 2017.

\bibitem{lin2016fixed}
D.~Lin, S.~Talathi, and S.~Annapureddy, ``Fixed point quantization of deep
  convolutional networks,'' in \emph{International conference on machine
  learning}, 2016, pp. 2849--2858.

\bibitem{wang2019haq}
K.~Wang, Z.~Liu, Y.~Lin, J.~Lin, and S.~Han, ``Haq: Hardware-aware automated
  quantization with mixed precision,'' in \emph{Proceedings of the IEEE
  conference on computer vision and pattern recognition}, 2019, pp. 8612--8620.

\bibitem{jacob2018quantization}
B.~Jacob, S.~Kligys, B.~Chen, M.~Zhu, M.~Tang, A.~Howard, H.~Adam, and
  D.~Kalenichenko, ``Quantization and training of neural networks for efficient
  integer-arithmetic-only inference,'' in \emph{Proceedings of the IEEE
  Conference on Computer Vision and Pattern Recognition}, 2018, pp. 2704--2713.

\bibitem{rusci2019memory}
M.~Rusci, A.~Capotondi, and L.~Benini, ``Memory-driven mixed low precision
  quantization for enabling deep network inference on microcontrollers,''
  \emph{arXiv preprint arXiv:1905.13082}, 2019.

\bibitem{lai2018cmsis}
L.~Lai, N.~Suda, and V.~Chandra, ``Cmsis-nn: Efficient neural network kernels
  for arm cortex-m cpus,'' \emph{arXiv preprint arXiv:1801.06601}, 2018.

\bibitem{garofalo2020pulp}
A.~Garofalo, M.~Rusci, F.~Conti, D.~Rossi, and L.~Benini, ``Pulp-nn:
  accelerating quantized neural networks on parallel ultra-low-power risc-v
  processors,'' \emph{Philosophical Transactions of the Royal Society A}, vol.
  378, no. 2164, p. 20190155, 2020.

\bibitem{bruschi2020enabling}
N.~Bruschi, A.~Garofalo, F.~Conti, G.~Tagliavini, and D.~Rossi, ``Enabling
  mixed-precision quantized neural networks in extreme-edge devices,'' in
  \emph{Proceedings of the 17th ACM International Conference on Computing
  Frontiers}, 2020, pp. 217--220.

\bibitem{garofalo2019pulp}
A.~Garofalo, M.~Rusci, F.~Conti, D.~Rossi, and L.~Benini, ``Pulp-nn: A
  computing library for quantized neural network inference at the edge on
  risc-v based parallel ultra low power clusters,'' in \emph{2019 26th IEEE
  International Conference on Electronics, Circuits and Systems (ICECS)}.\hskip
  1em plus 0.5em minus 0.4em\relax IEEE, 2019, pp. 33--36.

\bibitem{gautschi2017near}
M.~Gautschi, P.~D. Schiavone, A.~Traber, I.~Loi, A.~Pullini, D.~Rossi,
  E.~Flamand, F.~K. G{\"u}rkaynak, and L.~Benini, ``Near-threshold risc-v core
  with dsp extensions for scalable iot endpoint devices,'' \emph{IEEE
  Transactions on Very Large Scale Integration (VLSI) Systems}, vol.~25,
  no.~10, pp. 2700--2713, 2017.

\bibitem{M55}
D.~E. Joseph~Yiu, ``Introduction to the arm cortex-m55 processor. available
  online: https://pages.arm.com/cortex-m55-introduction.html,'' February 2020.

\bibitem{rusci2018work}
M.~Rusci, A.~Capotondi, F.~Conti, and L.~Benini, ``Work-in-progress: Quantized
  nns as the definitive solution for inference on low-power arm mcus?'' in
  \emph{2018 International Conference on Hardware/Software Codesign and System
  Synthesis (CODES+ ISSS)}.\hskip 1em plus 0.5em minus 0.4em\relax IEEE, 2018,
  pp. 1--2.

\bibitem{capotondi2020cmix}
A.~Capotondi, M.~Rusci, M.~Fariselli, and L.~Benini, ``Cmix-nn: Mixed
  low-precision cnn library for memory-constrained edge devices,'' \emph{IEEE
  Transactions on Circuits and Systems II: Express Briefs}, vol.~67, no.~5, pp.
  871--875, 2020.

\bibitem{cavigelli2016origami}
L.~Cavigelli and L.~Benini, ``Origami: A 803-gop/s/w convolutional network
  accelerator,'' \emph{IEEE Transactions on Circuits and Systems for Video
  Technology}, vol.~27, no.~11, pp. 2461--2475, 2016.

\bibitem{desoli201714}
G.~Desoli, N.~Chawla, T.~Boesch, S.-p. Singh, E.~Guidetti, F.~De~Ambroggi,
  T.~Majo, P.~Zambotti, M.~Ayodhyawasi, H.~Singh \emph{et~al.}, ``14.1 a 2.9
  tops/w deep convolutional neural network soc in fd-soi 28nm for intelligent
  embedded systems,'' in \emph{2017 IEEE International Solid-State Circuits
  Conference (ISSCC)}.\hskip 1em plus 0.5em minus 0.4em\relax IEEE, 2017, pp.
  238--239.

\bibitem{moons2017minimum}
B.~Moons, K.~Goetschalckx, N.~Van~Berckelaer, and M.~Verhelst, ``Minimum energy
  quantized neural networks,'' in \emph{2017 51st Asilomar Conference on
  Signals, Systems, and Computers}.\hskip 1em plus 0.5em minus 0.4em\relax
  IEEE, 2017, pp. 1921--1925.

\bibitem{lee2018unpu}
J.~Lee, C.~Kim, S.~Kang, D.~Shin, S.~Kim, and H.-J. Yoo, ``Unpu: A 50.6 tops/w
  unified deep neural network accelerator with 1b-to-16b fully-variable weight
  bit-precision,'' in \emph{2018 IEEE International Solid-State Circuits
  Conference-(ISSCC)}.\hskip 1em plus 0.5em minus 0.4em\relax IEEE, 2018, pp.
  218--220.

\bibitem{andri2017yodann}
R.~Andri, L.~Cavigelli, D.~Rossi, and L.~Benini, ``Yodann: An architecture for
  ultralow power binary-weight cnn acceleration,'' \emph{IEEE Transactions on
  Computer-Aided Design of Integrated Circuits and Systems}, vol.~37, no.~1,
  pp. 48--60, 2017.

\bibitem{moons201714}
B.~Moons, R.~Uytterhoeven, W.~Dehaene, and M.~Verhelst, ``14.5 envision: A
  0.26-to-10tops/w subword-parallel dynamic-voltage-accuracy-frequency-scalable
  convolutional neural network processor in 28nm fdsoi,'' in \emph{2017 IEEE
  International Solid-State Circuits Conference (ISSCC)}.\hskip 1em plus 0.5em
  minus 0.4em\relax IEEE, 2017, pp. 246--247.

\bibitem{gokhale2017snowflake}
V.~Gokhale, A.~Zaidy, A.~X.~M. Chang, and E.~Culurciello, ``Snowflake: An
  efficient hardware accelerator for convolutional neural networks,'' in
  \emph{2017 IEEE International Symposium on Circuits and Systems
  (ISCAS)}.\hskip 1em plus 0.5em minus 0.4em\relax IEEE, 2017, pp. 1--4.

\bibitem{ma2017automatic}
Y.~Ma, Y.~Cao, S.~Vrudhula, and J.-s. Seo, ``An automatic rtl compiler for
  high-throughput fpga implementation of diverse deep convolutional neural
  networks,'' in \emph{2017 27th International Conference on Field Programmable
  Logic and Applications (FPL)}.\hskip 1em plus 0.5em minus 0.4em\relax IEEE,
  2017, pp. 1--8.

\bibitem{meloni2018neuraghe}
P.~Meloni, A.~Capotondi, G.~Deriu, M.~Brian, F.~Conti, D.~Rossi, L.~Raffo, and
  L.~Benini, ``Neuraghe: Exploiting cpu-fpga synergies for efficient and
  flexible cnn inference acceleration on zynq socs,'' \emph{ACM Transactions on
  Reconfigurable Technology and Systems (TRETS)}, vol.~11, no.~3, pp. 1--24,
  2018.

\bibitem{qiu2016going}
J.~Qiu, J.~Wang, S.~Yao, K.~Guo, B.~Li, E.~Zhou, J.~Yu, T.~Tang, N.~Xu, S.~Song
  \emph{et~al.}, ``Going deeper with embedded fpga platform for convolutional
  neural network,'' in \emph{Proceedings of the 2016 ACM/SIGDA International
  Symposium on Field-Programmable Gate Arrays}, 2016, pp. 26--35.

\bibitem{prost2017scalable}
A.~Prost-Boucle, A.~Bourge, F.~P{\'e}trot, H.~Alemdar, N.~Caldwell, and
  V.~Leroy, ``Scalable high-performance architecture for convolutional ternary
  neural networks on fpga,'' in \emph{2017 27th International Conference on
  Field Programmable Logic and Applications (FPL)}.\hskip 1em plus 0.5em minus
  0.4em\relax IEEE, 2017, pp. 1--7.

\bibitem{umuroglu2017finn}
Y.~Umuroglu, N.~J. Fraser, G.~Gambardella, M.~Blott, P.~Leong, M.~Jahre, and
  K.~Vissers, ``Finn: A framework for fast, scalable binarized neural network
  inference,'' in \emph{Proceedings of the 2017 ACM/SIGDA International
  Symposium on Field-Programmable Gate Arrays}, 2017, pp. 65--74.

\bibitem{LatticeSENSEAI}
Lattice, ``{Lattice sensAI Delivers 10X Performance Boost for Low-Power, Smart
  IoT Devices at the Edge},''
  \url{https://www.latticesemi.com/About/Newsroom/PressReleases/2019/201911sensAI}.
  Last accessed on Sept. 20.

\bibitem{Raspberry}
``Raspberry pi compute module 3+. 2019,''
  \url{https://www.raspberrypi.org/documentation/hardware/computemodule/datasheets/rpi\_DATA\_CM3plus\_1p0.pdf}.

\bibitem{Kendryte}
``2018. kendryte: K210 datasheet,''
  \url{https://s3.cn-north-1.amazonaws.com.cn/dl.kendryte.com/documents/kendryte\_datasheet\_20181011163248\_en.pdf}.

\bibitem{movidius}
``Intel neural compute stick 2. high performance, low power for ai inference.''
  \url{https://www.intel.com/content/dam/support/us/en/documents/boardsandkits/neural-compute-sticks/NCS2_Product-Brief-English.pdf}.

\bibitem{trilium}
``Arm.project trillium machine learning platform,''
  \url{https://www.arm.com/products/silicon-ip-cpu/machine-learning/project-trillium}.

\bibitem{conti2015ultra}
F.~Conti and L.~Benini, ``A ultra-low-energy convolution engine for fast
  brain-inspired vision in multicore clusters,'' in \emph{2015 Design,
  Automation \& Test in Europe Conference \& Exhibition (DATE)}.\hskip 1em plus
  0.5em minus 0.4em\relax IEEE, 2015, pp. 683--688.

\bibitem{andri2020extending}
R.~Andri, T.~Henriksson, and L.~Benini, ``Extending the risc-v isa for
  efficient rnn-based 5g radio resource management,'' \emph{arXiv preprint
  arXiv:2002.12877}, 2020.

\bibitem{pullini2019mr}
A.~Pullini, D.~Rossi, I.~Loi, G.~Tagliavini, and L.~Benini, ``Mr. wolf: An
  energy-precision scalable parallel ultra low power soc for iot edge
  processing,'' \emph{IEEE Journal of Solid-State Circuits}, vol.~54, no.~7,
  pp. 1970--1981, 2019.

\bibitem{flamand2018gap}
E.~Flamand, D.~Rossi, F.~Conti, I.~Loi, A.~Pullini, F.~Rotenberg, and
  L.~Benini, ``Gap-8: A risc-v soc for ai at the edge of the iot,'' in
  \emph{2018 IEEE 29th International Conference on Application-specific
  Systems, Architectures and Processors (ASAP)}.\hskip 1em plus 0.5em minus
  0.4em\relax IEEE, 2018, pp. 1--4.

\bibitem{burrello2020dory}
A.~Burrello, A.~Garofalo, N.~Bruschi, G.~Tagliavini, D.~Rossi, and F.~Conti,
  ``Dory: Automatic end-to-end deployment of real-world dnns on low-cost iot
  mcus,'' \emph{arXiv preprint arXiv:2008.07127}, 2020.

\bibitem{wu2017parallel}
B.-C. Wu and I.-C. Wey, ``Parallel balanced-bit-serial design technique for
  ultra-low-voltage circuits with energy saving and area efficiency
  enhancement,'' \emph{IEEE Transactions on Circuits and Systems I: Regular
  Papers}, vol.~65, no.~1, pp. 141--153, 2017.

\bibitem{garofalo2020xpulpnn}
A.~Garofalo, G.~Tagliavini, F.~Conti, D.~Rossi, and L.~Benini, ``Xpulpnn:
  accelerating quantized neural networks on risc-v processors through isa
  extensions,'' in \emph{2020 Design, Automation \& Test in Europe Conference
  \& Exhibition (DATE)}.\hskip 1em plus 0.5em minus 0.4em\relax IEEE, 2020, pp.
  186--191.

\bibitem{choi2018pact}
J.~Choi, Z.~Wang, S.~Venkataramani, P.~I.-J. Chuang, V.~Srinivasan, and
  K.~Gopalakrishnan, ``Pact: Parameterized clipping activation for quantized
  neural networks,'' \emph{arXiv preprint arXiv:1805.06085}, 2018.

\end{thebibliography}
%
\vspace{-30pt}
\begin{IEEEbiography}[{\includegraphics[width=1in,height=1.25in,clip,keepaspectratio]{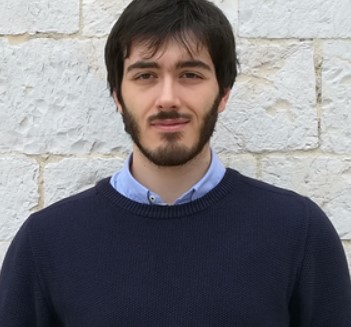}}]{Angelo Garofalo} received the B.Sc and M.Sc. degree in electronic engineering from the University of Bologna, Bologna, Italy, in 2016 and 2018 respectively. He is currently working toward his Ph.D. degree at DEI, University of Bologna, Bologna, Italy. His main research topic is Hardware-Software design of ultra-low power multiprocessor systems on chip. His research interests include Quantized Neural Networks, Hardware efficient Machine Learning, transprecision computing, and energy-efficient fully-programmable embedded architectures.
\end{IEEEbiography}
\vspace{-30pt}
\begin{IEEEbiography}[{\includegraphics[width=1in,height=1.25in,clip,keepaspectratio]{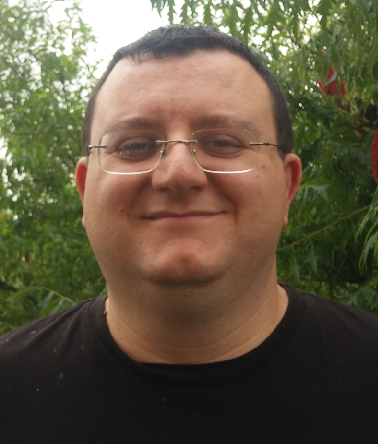}}]{Giuseppe Tagliavini}
received the Ph.D. degree in electronic engineering from the University of Bologna, Bologna, Italy, in 2017.
He is currently an Assistant Professor with the Department of Computer Science and Engineering (DISI) at the University of Bologna. He has co-authored over 30 papers in international conferences and journals. His research interests include parallel programming models for embedded systems, run-time optimization for multicore and many-core accelerators, and design of software stacks for emerging computing architectures.
\end{IEEEbiography}
\vspace{-30pt}
\begin{IEEEbiography}[{\includegraphics[width=1in,height=1.15in,clip,keepaspectratio]{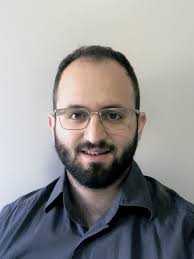}}]{Francesco Conti} received the Ph.D. degree in electronic engineering from the University of Bologna, Italy, in 2016. He is currently an Assistant Professor in the DEI Department of the University of Bologna. From 2016 to 2020, he held a research grant in the DEI department of University of Bologna and a position as postdoctoral researcher at the Integrated Systems Laboratory of ETH Zurich in the Digital Systems group.
His research focuses on the development of advanced deep learning based intelligence on top of ultra-low power, ultra-energy efficient programmable Systems-on-Chip -- from both the hardware and software perspective.
His research work has resulted in more than 40 publications in international conferences and journals and has been awarded several times, including the 2020 IEEE TCAS-I Darlington Best Paper Award.
\end{IEEEbiography}
\vspace{-30pt}
\begin{IEEEbiography}[{\includegraphics[width=1in,height=1.25in,clip,keepaspectratio]{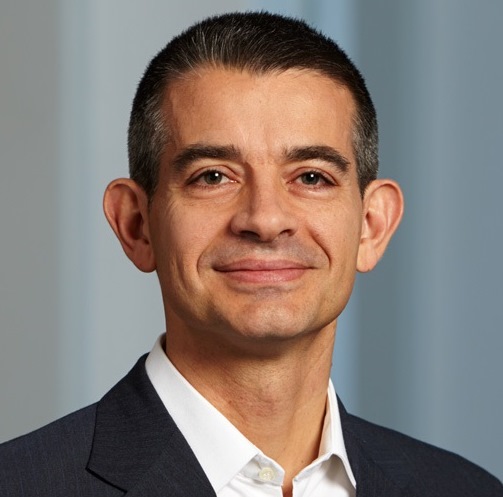}}]{Luca Benini}
holds the chair of digital Circuits and systems at ETHZ and is Full Professor at the Universita di Bologna.
Dr. Benini’s research interests are in energy-efficient computing systems design, from embedded to high-performance.
He has published more than 1000 peer-reviewed papers and five books.
He is a Fellow of the ACM and a member of the Academia Europaea.
He is the recipient of the 2016 IEEE CAS Mac Van Valkenburg Award and the 2020 EDAA Achievement Award.
\end{IEEEbiography}
\vspace{-30pt}
\begin{IEEEbiography}[{\includegraphics[width=1in,height=1.15in,clip,keepaspectratio]{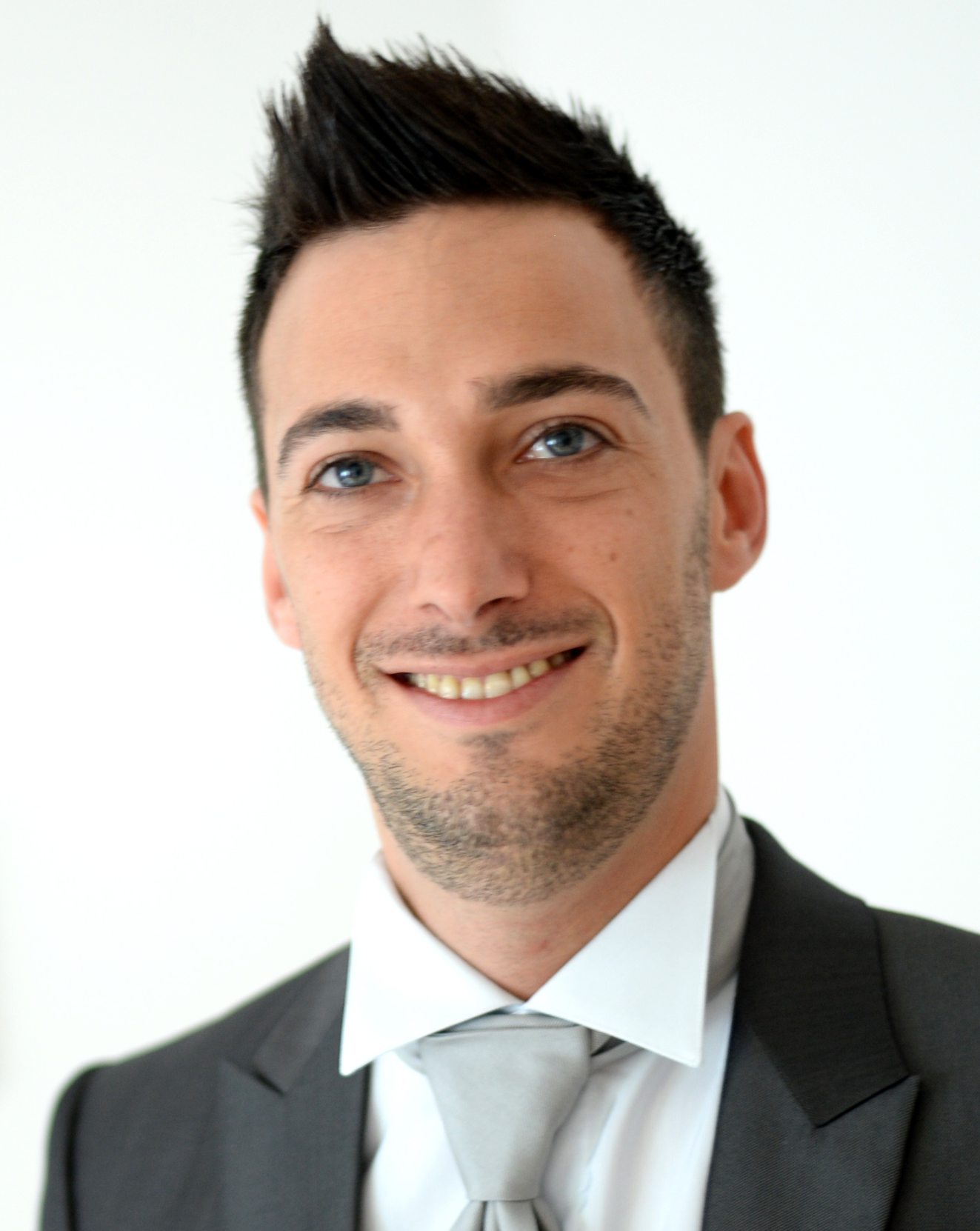}}]{Davide Rossi}, received the PhD from the University of Bologna, Italy, in 2012 where he currently holds an assistant professor position. His research interests focus on energy efficient digital architectures in the domain of heterogeneous and reconfigurable multi and many-core systems on a chip. This includes architectures, design implementation strategies, and runtime support to address performance, energy efficiency, and reliability issues of both high end embedded platforms and ultra-low-power computing platforms targeting the IoT domain. In these fields he has published more than 100 papers in international peer-reviewed conferences and journals. He is recipient of Donald O. Pederson Best Paper Award 2018, - 2020 IEEE Transactions on Circuits and Systems Darlington Best Paper Award, 2020 IEEE Transactions on Very Large Scale Integration Systems Prize Paper Award.
\end{IEEEbiography}
\vspace{-30pt}
%



\end{document}